\documentclass[11pt,a4paper]{article}
\pdfoutput=1

\usepackage{jcappub}
\usepackage[utf8]{inputenc}

\usepackage{amsmath}
\usepackage{amssymb} 
\usepackage[cm]{fullpage}
\usepackage{graphicx} 
\usepackage{siunitx}
\usepackage{tabularx} 
\usepackage[table,xcdraw]{xcolor}
\usepackage{fancyhdr}
\usepackage{braket}

\DeclareSIUnit\parsec{pc}
\numberwithin{equation}{section}

\newcommand{\beq}{\begin{equation}}
\newcommand{\eeq}{\end{equation}}

\newcommand{\rpar}[1]{\left(#1\right)}
\newcommand{\spar}[1]{\left[#1\right]}

\newcommand{\bd}{{\mathrm d}}

\newcommand{\eqtxt}[1]{\mathrel{\overset{\makebox[0pt]{\mbox{\normalfont\tiny\sffamily #1}}}{=}}}
\newcommand{\eqbg}{\eqtxt{\text{FRW}}}

\newcommand{\ga}{\gamma}

\newcommand{\phidot}{\dot{\phi}}
\newcommand{\varphidot}{\dot{\varphi}}
\newcommand{\chidot}{\dot{\chi}}
\newcommand{\thetadot}{\dot{\theta}}

\newcommand{\pp}{^{\phi \phi}}
\newcommand{\cc}{^{\chi \chi}}
\newcommand{\pc}{^{\phi \chi}}
\newcommand{\cp}{^{\chi \phi}}

\newcommand{\dpp}{_{\phi \phi}}

\newcommand{\dtt}{_{\theta \theta}}

\newcommand{\spp}{{<\phi \phi>}}
\newcommand{\scc}{{<\chi \chi>}}
\newcommand{\spc}{{<\phi \chi>}}
\newcommand{\scp}{{<\chi \phi>}}
\newcommand{\stt}{{<\theta \theta>}}

\newcommand{\cmodel}{^\text{C}}
\newcommand{\dmodel}{^\text{DBI}}
\newcommand{\dbi}{_d}
\newcommand{\gad}{\ga\dbi}

\newcommand{\cpoc}{\frac{C'}{C}}
\newcommand{\dpod}{\frac{D'}{D}}

\newcommand{\pperms}{+ \text{perms.}}
\newcommand{\TRS}{\mathcal{T}_\mathcal{RS}}

\newcommand{\tpcfrrr}{\braket{\mathcal{R}(\mathbf{k}_1)\mathcal{R}(\mathbf{k}_2)\mathcal{R}(\mathbf{k}_3)}}
\newcommand{\tpcf}[3]{\braket{Q^{#1}(\mathbf{k}_1)Q^{#2}(\mathbf{k}_2)Q^{#3}(\mathbf{k}_3)}}
\newcommand{\tpcfppp}{\tpcf{\phi}{\phi}{\phi}}
\newcommand{\tpcfpcc}{\tpcf{\phi}{\theta}{\theta}}
\newcommand{\tpcfppc}{\tpcf{\phi}{\phi}{\theta}}

\newcommand{\tpcfccc}{\tpcf{\theta}{\theta}{\theta}}

\newcommand{\ppcperms}{\braket{(Q^\phi)^2 Q^\theta}}
\newcommand{\pccperms}{\braket{Q^\phi (Q^\theta)^2}}

\newcommand{\doverc}{\frac{D}{C}}
\newcommand{\al}[1]{\alpha_{#1}}
\newcommand{\be}[1]{\beta_{#1}}

\newcommand{\rhochi}{\rho_\chi}
\newcommand{\pchi}{p_\chi}
\newcommand{\cppoc}{\frac{C''}{C}}
\newcommand{\dppod}{\frac{D''}{D}}
\newcommand{\gbp}[2]{\left(#1\ga^2 + #2\right)} 
\newcommand{\gbm}[2]{\left(#1\ga^2 - #2\right)} 

\newcommand{\phidotdot}{\ddot{\phi}}

\begin{document}

\title{Non-Gaussianity in multi-sound-speed disformally coupled inflation}

\author[a]{Carsten van de Bruck}
\author[b]{, Tomi Koivisto}
\author[a]{and Chris Longden}

\affiliation[a]{Consortium for Fundamental Physics, School of Mathematics and Statistics, University of Sheffield, Hounsfield Road, Sheffield S3 7RH, United Kingdom}
\affiliation[b]{Nordita, KTH Royal Institute of Technology and Stockholm University, 
Roslagstullsbacken 23, SE-10691 Stockholm, Sweden}

\abstract{Most, if not all, scalar-tensor theories are equivalent to General Relativity with a disformally coupled matter sector. In extra-dimensional theories such a coupling can be understood as a result of induction of the metric on a brane that matter is confined to. This article presents a first look at the non-Gaussianities in disformally coupled inflation, a simple two-field model that features a novel kinetic interaction. Cases with both canonical and Dirac-Born-Infeld (DBI) kinetic terms are taken into account, the latter motivated by the possible extra-dimensional origin of the disformality. The computations are carried out for the equilateral configuration in the slow-roll regime, wherein it is found that the non-Gaussianity is typically rather small and negative. This is despite the fact that the new kinetic interaction causes the perturbation modes to propagate with different sounds speeds, which may both significantly deviate from unity during inflation.} 

\maketitle

\section{Introduction}

Gravitation can be described by the dynamics of a metric $g_{\mu\nu}$ of spacetime interacting with matter fields. At the local limit of weak fields and slow motions, it has been well verified experimentally that the dynamics are given by the Einstein's field equations \cite{Will:2014kxa}, whose predictions have now been confirmed also in the nonlinear regime by the data from the recent observation of gravitational waves generated during the merging of two black holes \cite{Abbott:2016blz}. On cosmological scales, however, the observational data clearly forces us to modify either the matter sources for the $g_{\mu\nu}$ or the dynamics of the interaction \cite{Clifton:2011jh}.
\newline
\newline
A basic way to introduce such modifications is to let matter couple to a metric $\hat{g}_{\mu\nu}$ that may be distinct from the spacetime metric $g_{\mu\nu}$. The case of a conformal  
relation, $\hat{g}_{\mu\nu}=Cg_{\mu\nu}$, corresponding to a rescaling of units, arises quite typically in generalised theories, for example, from the coupling of the dilaton in the low-energy limit of string theory, in which case the transformation is given in terms of a scalar field $\phi$ as $C(\phi)=\phi^2$. However, the most general form of a relation $\hat{g}_{\mu\nu}=\hat{g}_{\mu\nu}(\phi,g_{\mu\nu})$ which still complies with physical requirements such as causality, is a so called disformal relation, \cite{Bekenstein:1992pj}

\beq \label{eq:DisformalMetric}
\hat{g}_{\mu \nu} = C(\phi,X) g_{\mu \nu} + D(\phi,X) \phi_{,\mu} \phi_{,\nu} \,,
\eeq

that may involve also the derivatives $\phi_{,\mu}$ of the scalar field, and even nonlinear dependence upon the metric through the kinetic term $X = -g^{\mu\nu} \phi_{,\mu} \phi_{,\nu} / 2$.  
Their extended range of phenomenology \cite{Kaloper:2003yf, Zumalacarregui:2010wj,Sakstein:2015jca,Sakstein:2014aca,Emond:2015efw,Burrage:2016myt,Brax:2015hma,Minamitsuji:2016hkk,Bettoni:2016mij} and universality in scalar-tensor theories \cite{Zumalacarregui:2012us,Zumalacarregui:2013pma,Bettoni:2013diz,Arroja:2015wpa,Achour:2016rkg}
have made disformal transformations rather popular in cosmology and modified gravity in recent years \cite{vandeBruck:2012vq,vandeBruck:2015ida,Motohashi:2015pra,Minamitsuji:2014waa,Domenech:2015hka,Hagala:2015paa} (whilst they have also found novel applications in theory \cite{Yuan:2015tta,Bittencourt:2015ypa,Huang:2015hja,Carvalho:2015omv,Bittencourt:2016smd}).
\newline
\newline
Disformally coupled inflation \cite{vandeBruck:2015tna} was introduced as a model of inflation with two scalar fields, $\phi$ and $\chi$, which both minimally couple to gravity, but on different metrics, such that the action is written as

\begin{align}
S= \frac{1}{2} \int {\rm d}^4 x \sqrt{-g} \ R \ - \ \int {\rm d}^4 x \sqrt{-g}\left[\frac{1}{2} g^{\mu\nu}\phi_{,\mu}\phi_{,\nu} + U(\phi)\right] \ - \ \int {\rm d}^4 x \sqrt{-\hat{g}}\left[ \frac{1}{ 2}\hat{g}^{\mu\nu}\chi_{,\mu}\chi_{,\nu} + V(\chi)\right] \,, \label{eq:DisformalAction}
\end{align}

where the two metrics $g$ and $\hat{g}$ are related by a transformation of the form

\beq 
\hat{g}_{\mu \nu} = C(\phi) g_{\mu \nu} + D(\phi) \phi_{,\mu} \phi_{,\nu} \,. \label{eq:Dmetric}
\eeq

The action (\ref{eq:DisformalAction}) can be regarded as a phenomenological parameterisation of a minimal disformal setting, where the matter sector consists solely of a canonical scalar field $\chi$, and the coupling functions are taken to depend on $\phi$ and not the kinetic term $X$ for simplicity. The model has then four arbitrary functions, the potentials $U(\phi)$ and $V(\chi)$ as well as the conformal and disformal factors $C(\phi)$ and $D(\phi)$.
\newline
\newline
A new feature arising from the coupling $D$ is the so called kinetic mixing \cite{Bettoni:2015wta} of the scalar fields. Using the relation (\ref{eq:Dmetric}) to rewrite the above action in terms of only the metric $g$, it is transformed into a two-field scalar-tensor theory with non-trivial derivative interactions,

\beq \label{eq:CanonicalModel}
S = \int {\bd}^4 x \sqrt{-g}\left[\frac{1}{2} R + X\pp - U(\phi) + \frac{C(\phi)}{\ga}\left[X\cc - C(\phi) V(\chi) \right] + 2 \ga D(\phi) (X\pc)^2 \right] \,,
\eeq

where we have introduced the notation for kinetic terms $X^{IJ} = - 1/2 \partial^\mu \phi^I \partial_\mu \phi^J$, defined on the metric $g$, where $I$ and $J$ are field indices that can take values from $(\phi, \chi)$.  Also appearing in this action is the parameter $\ga$, which is defined by

\beq
\ga = \frac{1}{\sqrt{1 - 2 \frac{D}{C} X\pp}} \,. \label{gamma}
\eeq

Its presence in the action may appear dangerous, as the argument of the square root is not in general positive definite. The $\gamma$-factor can become infinite when the relation (\ref{eq:Dmetric}) is non-invertible. However, this situation is analogous to the conformally coupled models where the function $C$ may in principle vanish as a function of the field, but this does not occur in the dynamics of physical models. In cosmological backgrounds, where $2X\pp = \dot{\phi}^2$, the $\gamma$-factor resembles the Lorentz boost in special relativity, due to which it would take an infinite time to reach the singularity of the transformation from the point of view of a physical observer. This feature has been verified in explicit calculations in both the Jordan and Einstein frames, see e.g. \cite{Zumalacarregui:2012us,Sakstein:2015jca}. Not coincidentally, the form of $\gamma$ is familiar from the Dirac-Born-Infeld (DBI) brane inflation, where it determines the ''cosmic speed limits'' \cite{Silverstein:2003hf,Alishahiha:2004eh}, if we identify the field $\phi$ as (proportional to) the radial coordinate of the D3-brane whose warp factor (squared) is given by $h(\phi) = D(\phi)/C(\phi)$.
\newline
\newline
This is due to the fact that in brane world models such as the DBI inflation \cite{Silverstein:2003hf,Alishahiha:2004eh}, the induced metric on the brane is generically disformal with respect to the bulk metric. The radial coordinate $\phi$ can acquire a kinetic term from a mere cosmological constant in the $\hat{g}$-frame, which transformed into the $g$-frame becomes $\sim h^{-2}/\gamma$. Further, in the case of maximal supersymmetry in type IIb compactification, the D3-brane is BPS and one obtains 
the DBI kinetic term $\sim h^{-2}(1/\gamma-1)$ (the latter correction could also be absorbed in the potential $V(\phi)$, that can originate from non-perturbative effects) \cite{Baumann:2010sx}. 
In an embedding of our model in such an extra-dimensional set-up, the $\chi$-field can be interpreted as matter that is confined to the brane, as in several previous cosmological scenarios exploiting brane matter living in the induced (and thus disformal) metric \cite{Cembranos:2003mr,Koivisto:2013fta,Koivisto:2014gia,Cembranos:2016jun}. 
More properly then, we would adopt the DBI kinetic term for the field $\phi$, and consider the action\footnote{Note that the parameter $\gamma$ also defines the DBI kinetic term, as both the kinetic interactions between the two fields and the DBI behaviour of $\phi$ both originate from the disformal relation between metrics; the DBI warp factor $h(\phi) = D(\phi)/C(\phi)$ is also entirely specified by the choice of conformal and disformal factors. The DBI kinetic term hence does not introduce any further arbitrary functions to the model. While we could consider DBI kinetic terms which are specified by a warp factor taken to be unrelated to the disformal coupling, in the interest of simplicity and physical motivation we do not explore this avenue.} 

\beq \label{eq:DBIModel}
S = \int {\bd}^4 x \sqrt{-g}\left[\frac{1}{2} R + \frac{C(\phi)}{D(\phi)}\rpar{1 - \frac{1}{\ga}} - U(\phi) + \frac{C(\phi)}{\ga}\left(X\cc - C(\phi) V(\chi) \right) + 2 \ga D(\phi) (X\pc)^2 \right] \,.
\eeq

In our previous work \cite{vandeBruck:2015tna}, we neglected the effects due to the DBI term to simplify our analysis and focus on effects due to the disformal interactions of the two fields, but in this work we shall reinstate it and consider its effects on inflation and non-Gaussianity too. We shall henceforth refer to eq. (\ref{eq:CanonicalModel}) as the canonical model of disformally coupled inflation, and eq. (\ref{eq:DBIModel}) as the DBI model. As a sanity check, we also note that when the kinetic term $X\pp$ is small and/or disformal coupling is weak compared to conformal coupling such that overall $D X\pp \ll C$, we can expand $\ga$ as $1 + D/C X\pp$ which upon substitution into the action reveals, as expected, that the DBI model reduces to the canonical model in this limit.
\newline
\newline
Considering the vast number of inflationary models that have been conceived since the origin of the paradigm, it is important to understand how well each model agrees with existing experimental data \cite{Martin:2013tda,Escudero2016}, usually via computation of the primordial power spectrum of curvature fluctuations. The most precise constraints currently available are those from CMB anisotropy probes, particularly those from the recent Planck mission \cite{Ade:2015lrj}. Great progress has been made in such endeavours to the point where fiducial power-law potential models of single field inflation are now essentially ruled out by their overly-large predictions of the tensor-to-scalar-ratio \cite{Ade:2015tva}. It is likely that the next generation of CMB experiments will shed further light on the nature of inflation through even tighter constraints on the parameters of the primordial power spectrum, but also, the first high-precision measurements of the deviation from purely Gaussian statistics - the non-Gaussianity - present in primordial fluctuations. In anticipation of non-Gaussianity becoming a powerful discriminator of inflationary models in the future, many cosmologists are already giving considerable thought to its computation and analysis in a wide range of models \cite{Seery:2005wm,Peterson2011,Maldacena:2002vr,Langlois:2008qf,Arroja:2008yy,Palma:2015eth}. For more comprehensive review articles, see \cite{Bartolo:2004if,Byrnes:2010em,Chen:2010xka,Chen:2006nt}. 
\newline
\newline
In this article, we extend our previous work on disformally coupled inflation to investigate the non-Gaussianity it predicts, both for the canonical and DBI models presented above. A particularly interesting aspect of the models is that perturbations in $\phi$ and $\chi$ each propagate with a different speed, which is a rarely-studied possibility in multi-field inflationary models and may influence non-Gaussianity in a novel fashion. Previous work on multi-field generalisations of the DBI models include various scenarios with additional moduli \cite{Huang:2007hh,Langlois:2008qf} but also the ''multi-speed inflation'' with additional DBI branes \cite{Cai:2008if,Cai:2009hw}. The action for the latter consists of a sum of DBI actions with different parameters, so that one has several decoupled k-essence fields with each an evolving sound speed of its own, with indeed interesting consequences to the structure of the generated non-Gaussianity \cite{Cai:2008if,Cai:2009hw,Emery:2012sm,Emery:2013yua,Pi:2011tv}. In contrast to these models, the disformally coupled inflation features a derivative interaction of the fields that is already apparent in the action.
\newline
\newline
The main body of the article proceeds as follows. In section \ref{sec:DBIspec} we repeat the analysis of our previous work on the canonical model for the DBI variant of the model introduced above to obtain its inflationary predictions, and find some example trajectories which fit the data. Using our previous results and this, in section \ref{sec:NGcalculations} we then turn our attention to non-Gaussianity and compute a leading-order estimate of the parameter $f_{NL}$ for the feasible trajectories found for both canonical and DBI models. In section \ref{sec:conclusions} we summarize our findings. In the appendix we provide the expressions for coefficients appearing in the perturbation equations and the derivatives of the Lagrangian with respect to the kinetic terms. 

\section{Power spectra with DBI kinetic term} \label{sec:DBIspec}

We proceed to repeat the analysis of our previous work on the canonical model for the full DBI model, culminating in a full numerical integration of the first order perturbations of the fields to obtain power spectra exactly, following the methods presented in \cite{Tsujikawa:2003ccc}. In the derivations, we shall take advantage of the formalism developed for 
generalised multi-field actions \cite{Langlois:2008qf,Langlois:2008mn,Arroja:2008yy} of the form

\beq \label{eq:GeneralLagrangian}
S = \int \bd^4 x \sqrt{-g} \spar{\frac{1}{2} R + P(\phi^I,X^{JK})} \,,
\eeq

where $(I,J,K = 1 \ldots N)$ for $N$ fields, and again, kinetic terms are denoted $X^{JK} = - 1/2 \partial^\mu \phi^J \partial_\mu \phi^K$. Because the action is not restricted to depending on single-field kinetic terms like $X\pp$ or $X\cc$, but also allowed to depend on mixed-kinetic terms like $X\pc$, even models with non-trivial kinetic interactions can be described in this way. 
In particular, both our canonical (\ref{eq:CanonicalModel})) and DBI (\ref{eq:DBIModel}) versions of the disformally coupled inflation are within the scope of (\ref{eq:GeneralLagrangian}).
As the kinetic terms obey the symmetry $X^{JK} = X^{KJ}$, there are $N(N+1)/2$ unique kinetic terms that the action is allowed to depend on. For this reason it is also convenient to define and use a symmetrised derivative with respect to kinetic combinations, 
\beq \label{eq:symderiv}
P_{<JK>} = \frac{1}{2} \rpar{\frac{\partial P}{\partial X^{JK}} + \frac{\partial P}{\partial X^{KJ}}} \,.
\eeq
The energy-momentum tensor for the general class of multi-field theories given by eq. (\ref{eq:GeneralLagrangian}) is given by
\beq
T^{\mu \nu} = P g^{\mu \nu} + P_{<IJ>} \partial^\mu \phi^I \partial^\nu \phi^J \,. \label{set}
\eeq
We can then derive the equations of motion for the inflationary background and the field fluctuations about it.

\subsection{Background equations}

We assume a flat universe throughout this article, so that the metric reads 
\beq
ds^2 = -dt^2 + a(t)^2dx_i dx^i\,.
\eeq
Using (\ref{set}), we can then readily obtain the Friedmann equations ($H = \dot a/a$),
\begin{align}
3H^2 = \rho & = 2 P_{<IJ>}X^{IJ} - P \, ,\\
2\dot{H} & = -(\rho+P) =  -2 P_{<IJ>}X^{IJ} \, .
\end{align}
Specialising this result to the DBI model Lagrangian in eq. (\ref{eq:DBIModel}), we find the energy density and pressure in this theory to be,
\begin{align}
\rho =\frac{C}{D}\rpar{\ga - 1} + U + \ga C \rpar{\ga^2 X\cc + CV} \, , \\
P = \frac{C}{D}\rpar{1 - \frac{1}{\ga}} - U + \frac{C}{\ga} \rpar{\ga^2 X\cc - CV} \, ,
\end{align}
where $\ga$ takes the form
\beq \label{eq:FRWgamma}
\ga = \frac{1}{\sqrt{1 - \frac{D}{C} \dot{\phi}^2}} \, .
\eeq
We can identify in both of these results the non-standard kinetic and potential terms associated with the $\chi$ field, 
\begin{align}
\rho_\chi & = \ga C \rpar{\ga^2 X\cc + CV} \, , \\
P_\chi & = \frac{C}{\ga} \rpar{\ga^2 X\cc - CV} \, ,
\end{align}
which are, of course, unchanged by the presence of a DBI kinetic term in the $\phi$ field. 
Similarly, the field equations of motion are given by,
\beq
K_{IJ} \ddot{\phi}^J + 3 H P_{<IJ>}\dot{\phi}^J  + 2 P_{<IJ>,K} X^{KJ} -  P_{,I} = 0 \, ,
\eeq
where the kinetic matrix $K_{IJ}$ is defined by,
\beq \label{eq:KineticMatrix}
K_{IJ} = P_{<IJ>} + 2 P_{<MJ><IK>}X^{MK} \, .
\eeq
Evaluating this explicitly for $I = \chi$ gives,
\beq
\ddot{\chi} + 3H \chidot + \ga^2 \frac{D}{C} \phidot \chidot \ddot{\phi} - \frac{1}{2}\spar{(\ga^2 - 3) \cpoc - (\ga^2 -1) \dpod}\phidot \chidot +  \frac{C}{\ga^2} V' = 0 \, ,
\eeq
which is in agreement with the result for the canonical case, as expected.  The $\phi$ equation of motion is similarly found to be,

\begin{align}
\rpar{\ga^3+ \ga^2 \frac{D}{C} \rho_\chi} \ddot{\phi} + 3 H \phidot \rpar{\ga - \ga^2 \frac{D}{C} p_\chi} & + U' +\frac{1}{2} (\ga^2-1)\rho_\chi \dpod - \frac{1}{2}\spar{(\ga^2-2)\rho_\chi + 3 \ga^2 p_\chi}\cpoc \nonumber \\ 
& + \frac{1}{2} \frac{C}{D} \rpar{\dpod-\cpoc} \rpar{\ga^3 - 3 \ga + 2}= 0\,,
\end{align}

which is manifestly different to the canonical equation of motion in that the kinetic terms proportional to $\ddot{\phi}$ and $\dot{\phi}$ are modified, and an extra term which appears in all theories with a DBI kinetic term has of course been generated. 

\subsection{Perturbation equations}

After a tedious but straightforward process, the perturbed field equations are found to be of the form

\begin{align}
2 \left(\partial_i \partial^i \Psi - 3 H \dot{\Psi}\right)  = & \ \delta\rho = X_1 \Psi + X_2 \delta\phi + X_3 \dot{\delta\phi} + X_4 \delta\chi + X_5 \dot{\delta\chi} \, \label{eq:PEEtt},
\\ \nonumber \\
2 \left(\dot{\Psi} + H \Psi \right) =& \ -\delta q  =  - Y_1 \delta\phi + -Y_2 \delta \chi \,\label{eq:PEEts} ,
\\ \nonumber \\
2 \left(\ddot{\Psi} + 4H\dot{\Psi} + 4\dot{H}\Psi + 6H^2 \Psi \right) = & \ \delta p  = Z_1 \Psi + Z_2 \delta\phi + Z_3 \dot{\delta\phi} + Z_4 \delta\chi +Z_5 \dot{\delta\chi}\label{eq:PEEss} \, ,
\end{align}

with $X_n$, $Y_n$ and $Z_n$ defined in appendix \ref{App:XYZ} and $\Psi$ being the metric perturbation in Newtonian gauge, such that metric perturbations about the cosmological background are given by,

\beq
\delta g_{\mu \nu} = -2 \ \text{diag} (\Phi, a^2(t) \Psi, a^2(t) \Psi, a^2(t) \Psi) \, ,
\eeq

and $\Phi = \Psi$ due to the off-diagonal spatial elements of the perturbed field equations. Note that these equations are of the same general structure as those previously obtained for the canonical case, depending on the same perturbed variables but with different coefficient functions.
\newline
\newline
We also obtain equations of motion for the field fluctuations, that is the perturbed (generalised) Klein-Gordon equations, in the form

\begin{align}
\al{1} \delta \ddot{\phi} + \al{2} \delta \ddot{\chi} + \al{3}\partial_i \partial^i \delta\phi + \al{4}\partial_i \partial^i \delta\chi + \al{5} \dot{\Psi} + \al{6} \delta \dot{\phi} + \al{7} \delta \dot{\chi} + \al{8} \Psi + \al{9} \delta \phi + \al{10} \delta \chi  = 0 \, ,  \label{eq:alphas} 
\\ \nonumber \\
\be{1} \delta \ddot{\phi} + \be{2} \delta \ddot{\chi} + \be{3}\partial_i \partial^i \delta\phi + \be{4}\partial_i \partial^i \delta\chi + \be{5} \dot{\Psi} + \be{6} \delta \dot{\phi} + \be{7} \delta \dot{\chi} + \be{8} \Psi + \be{9} \delta \phi + \be{10} \delta \chi   = 0 \, , \label{eq:betas} 
\end{align}

where the coefficients $\al{n}$ and $\be{n}$ are explicitly given in appendix \ref{App:AlphaDBI}. Compared to the canonical case previously studied, once again the structure of the equations of motion are unchanged and in particular, only the $\alpha_n$ coefficients are modified by the presence of the DBI kinetic term. From this, we can proceed to use the perturbed field equations to eliminate the gauge variable $\Psi$ and its first derivative with no significant difference to the canonical model except the differences in $X_n$, $Y_n$ and $Z_n$ coefficients, and numerically solve the resulting system written in terms of Sasaki-Mukhanov variables.  As in the previous work, we identify the combination of Sasaki-Mukhanov variables\footnote{Note that this expression was mistyped in eq. (3.4) of our previous work, missing the factor $\dot{\phi}\dot{\chi}$, though the correct expression was used in all numerical work. The expression presented here for $Q_\theta$ is corrected. This error in the previous manuscript propagated into its eq. (3.16), whose correct form is given later in this paper by eq. (\ref{eq:CorrectedR}). }

\beq \label{eq:Qthetadef}
Q^\theta = Q^\chi + \ga^2 \frac{D}{C} \phidot \chidot Q^\phi \, ,
\eeq

to be a more natural choice of variable than $Q_\chi$ as it diagonalises the kinetic and gradient terms in the second order action. Using this diagonal basis, it is simple to obtain the propagation speeds of the two fields, as

\begin{align}
c_s^{(\theta)} = \frac{1}{\ga} \, , \label{eq:DBIthetasoundspeed} \\
c_s^{(\phi)} = \sqrt{\frac{C - \ga D p_\chi}{\ga^2 C + \ga D \rho_\chi}} \, . \label{eq:DBIphisoundspeed}
\end{align}

Once again, the $\theta$ propagation speed is unchanged from the canonical case, but we find extra factors of $\ga$ appearing in the expression for the DBI-field kinetic term. This is expected, as if one ignores the terms pertaining to the second field in the above expression, $c_s^{(\phi)}$ reduces to $1/\ga$, the usual sound speed for a DBI field.
\newline
\newline
We can also obtain the curvature perturbation in our model,

\begin{align}
\mathcal{R} & = \rpar{\frac{H}{2 P_{<IJ>} X^{IJ}}}P_{<KL>} \dot{\phi}^K Q^L \nonumber \, \\
& = \rpar{\frac{H}{\rho + p}}\spar{ \rpar{\ga + \frac{D}{C} \rho_\chi} \phidot Q^\phi + \ga C \chidot Q^\chi} \nonumber \, \\
& = \rpar{\frac{H}{\rho + p}}\spar{ \rpar{\ga - \ga^2 \frac{D}{C} p_\chi} \phidot Q^\phi + \ga C \chidot Q^\theta} \label{eq:CorrectedR} \, .
\end{align}

Finally, we note that as all of the derivations in this section had already been done for the canonical model in our previous work, but via different methods not relying on results from the literature on general $P(\phi^I,X^{JK})$ theories, we took this opportunity to rederive all our results from the canonical model using these new methods with the Lagrangian in eq. (\ref{eq:CanonicalModel}) to confirm their consistency with our previous approach. We found that the two methods produced concordant results.

\subsection{Results}

As with the canonical model, we find that the DBI model of disformally coupled inflation is capable of making inflationary predictions in line with observations. Particularly, for fiducial massive field potentials,

\beq
U(\phi) = \frac{1}{2} m_\phi^2 \phi^2 \, , \quad V(\phi) = \frac{1}{2} m_\chi^2 \chi^2 \, , \label{pots}
\eeq

and exponential conformal and disformal factors,

\beq
C(\phi) = C_0 e^{c \phi} \, , \quad D(\phi) = D_0 e^{d \phi} \, , \label{cots}
\eeq

we observe that it is possible to achieve non-standard slow-roll inflationary trajectories where the disformal warping, parametrised by $\ga$ as given in eq. (\ref{eq:FRWgamma}), deviates from unity (the case when $D = 0$). Note that this is dependent on our phenomenologically-motivated choice of conformal and disformal coupling functions\footnote{In particular, the inclusion of the DBI kinetic term does not change the arguments in our previous work regarding the insipidity of the case $C \propto 1/D \propto \phi^2$ that corresponds to an adS-warp in the extra dimensional geometry. This is expected since we restrict to the slow-roll regime in the present study. The adS case being the most extensively studied in the type IIb settings (and in particular, a good approximation to the solution found by Klebanov and Strassler \cite{Klebanov:2000hb}), one is tempted to study its possible relevance to the disformally coupled DBI in the relativistic regime (meaning fast-roll in the sense $\gamma \gg 1$), but this is outside the scope of the paper at hand.}.

\begin{figure}
    \centering
    \includegraphics[width=\textwidth]{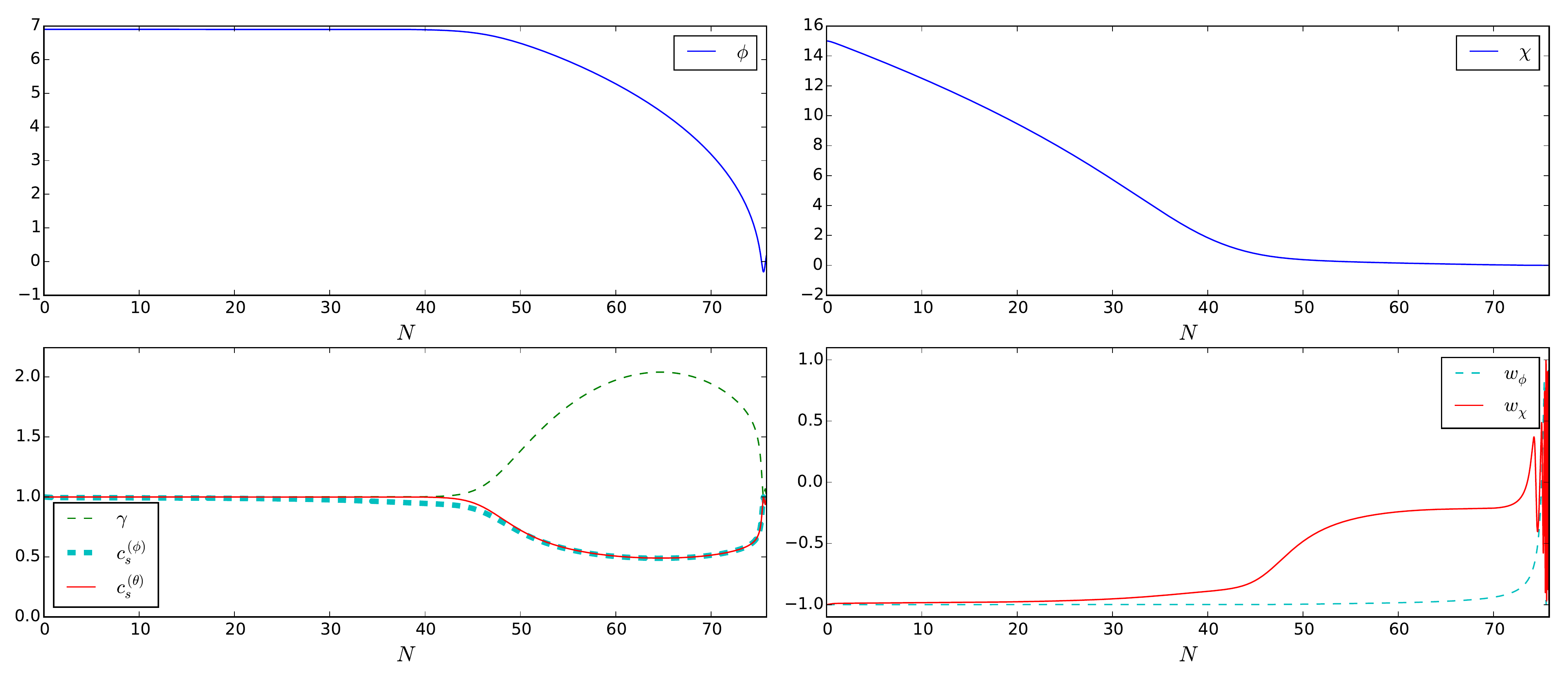}
    \includegraphics[width=\textwidth]{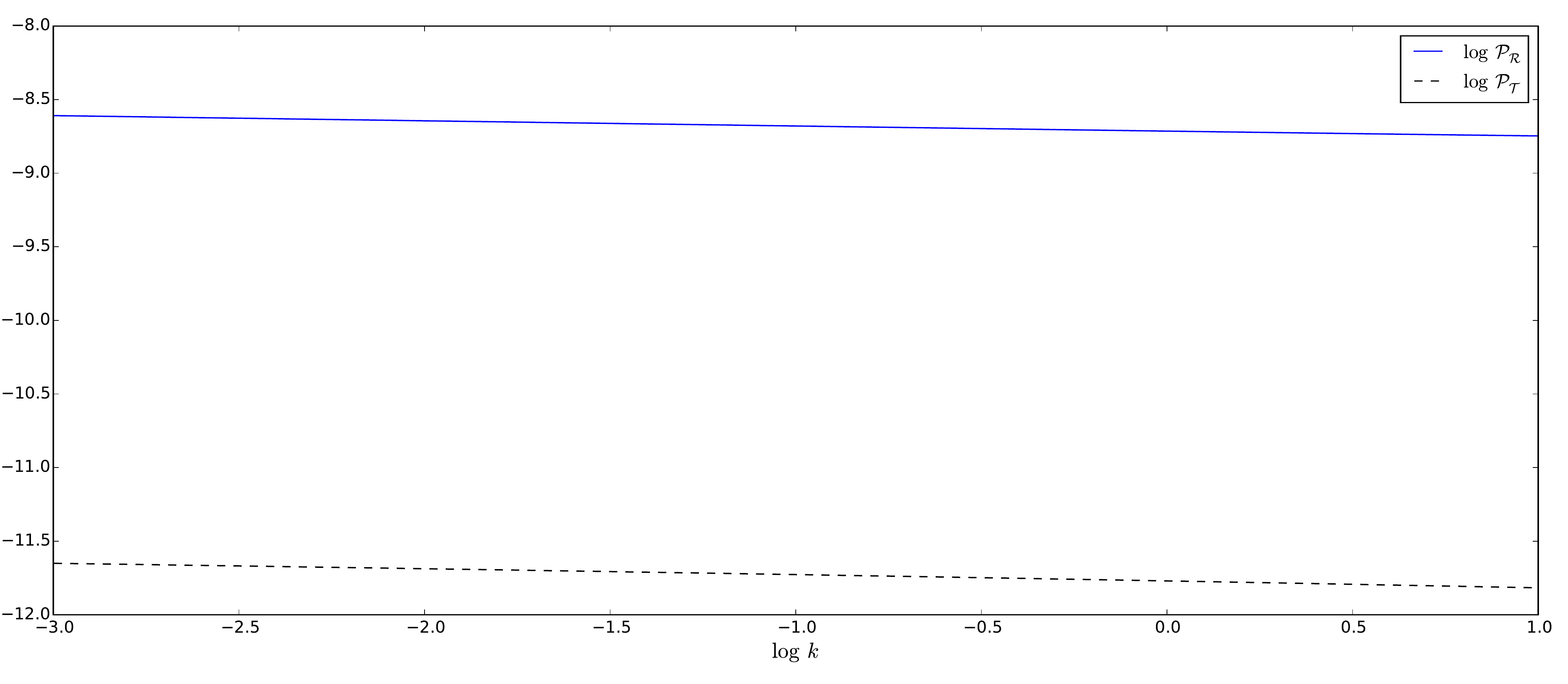}
    \caption{Trajectory A : Inflationary dynamics for the action (\ref{eq:DBIModel}) with the potentials (\ref{pots}) and couplings (\ref{cots}). Parameters used are $m_\chi = 2.9 m_\phi = 2.286 \times 10^{-6}$, $C_0 = 1$,$D_0 = 3.8 \times 10^{12}$, $d = -c = 0.1$ and initial conditions are $\phi_0 = 6.9$, $\chi_0 = 15.0$.}
     \label{fig:DBIA}
\end{figure}

\begin{figure}
    \centering
    \includegraphics[width=\textwidth]{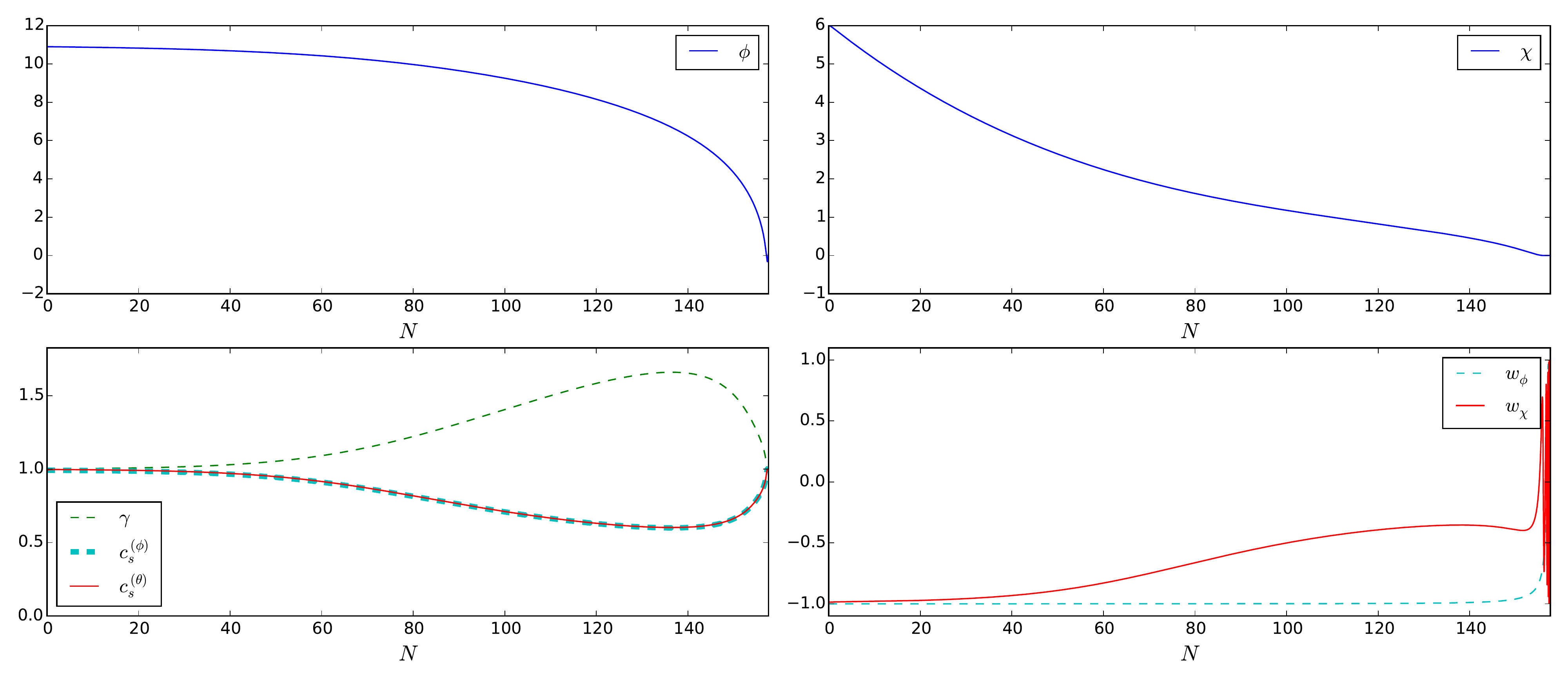}
    \includegraphics[width=\textwidth]{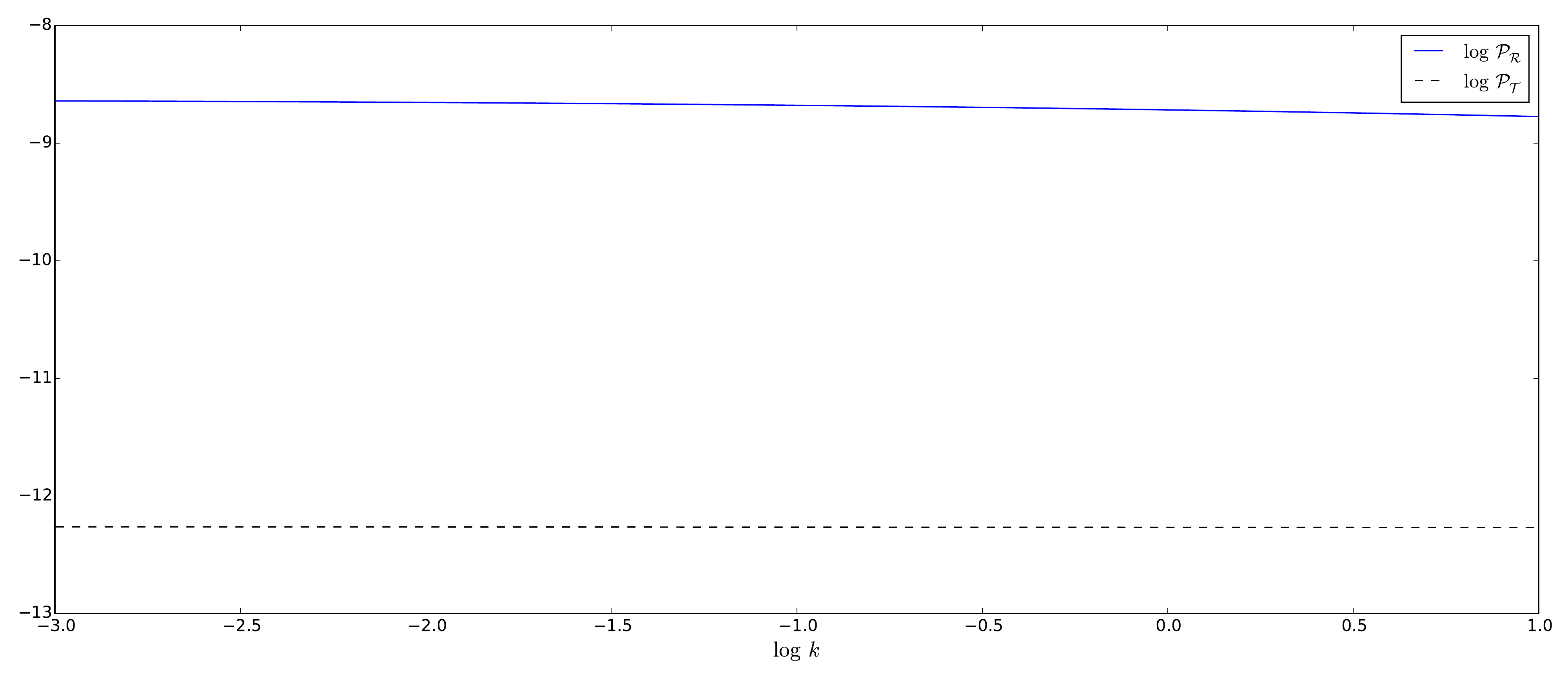}
    \caption{Trajectory B : Inflationary dynamics for the action (\ref{eq:DBIModel}) with the potentials (\ref{pots}) and couplings (\ref{cots}). Parameters used are $m_\chi = 3.1 m_\phi = 2.25 \times 10^{-6}$, $C_0 = 1$,$D_0 = 7.0 \times 10^{11}$, $d = -c = 0.21$ and initial conditions are $\phi_0 = 10.9$, $\chi_0 = 6.0$.}
    \label{fig:DBIB}
\end{figure}

Disformal effects during inflation lead to varying sound speeds and equations of state for the two fields, which translates into modifications to the power spectrum. In particular, we again find that a key result is boosting of the scalar amplitude due to subluminal propagation speeds, and in turn a small tensor-to-scalar-ratio. Examples of this are shown in figures \ref{fig:DBIA} (Trajectory A) and \ref{fig:DBIB} (Trajectory B), with $r$ values of $2 \times 10^{-3}$ and $7 \times 10^{-3}$ respectively. In contrast to the trajectories we previously obtained for the canonical model, however, both of these examples show that initially $\ga$ remains small and grows larger at late times, before eventually returning to around unity at the end of inflation. In Trajectory B, the small-$\ga$ phase occurs more than 100 e-folds before the end of inflation, and so is not relevant when observables modes are leaving the horizon, but in Trajectory A we present a more extreme case where during horizon crossing, $\ga \approx 1$ and it is only after horizon exit that disformal effects become large.

\begin{figure}
    \centering
    \includegraphics[width=\textwidth]{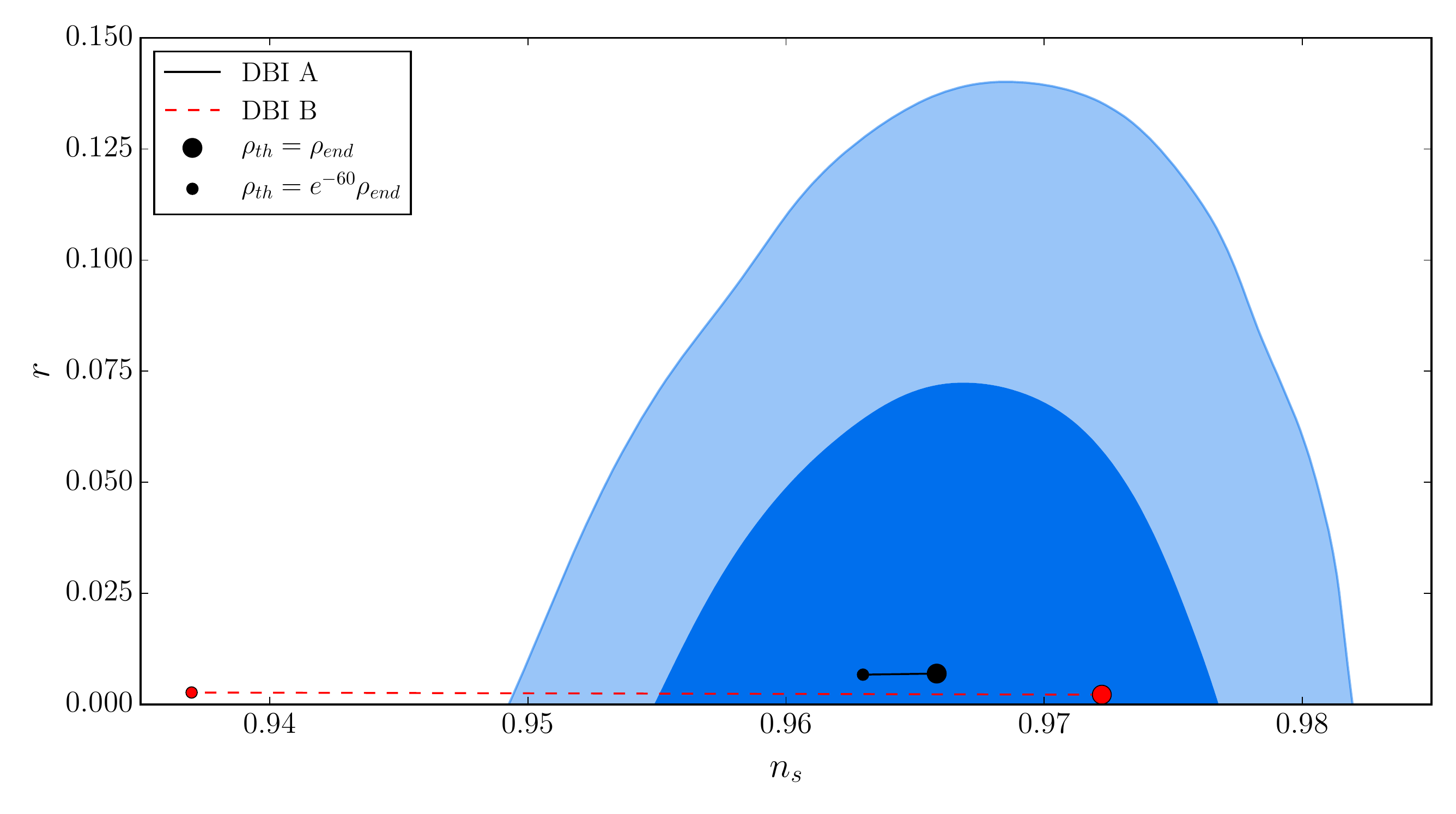}
    \caption{Predictions for $n_s$ and $r$ of Trajectories A and B for two different reheating scenarios, compared to the Planck 1$\sigma$ and $2\sigma$ contours. In the first scenario (large dots) the energy density at reheating is comparable to that at the end of inflation, representing an efficient reheating mechanism, while in the second scenario (small dots) it is much lower, such that the observable window is shifted by 5 e-folds compared to the first scenario. Trajectory A is fairly stable under the details of the reheating process, while Trajectory B requires somewhat efficient reheating to remain viable. }
     \label{fig:nsr}
\end{figure}

These two models are plotted in the $n_s$-$r$ plane in figure \ref{fig:nsr}. Here, model predictions are shown for two cases in which the details of reheating differ, shifting the location of the observable window. The amount of e-folds by which reheating affects this is given by \cite{Ade:2015lrj},

\beq
\Delta N = \frac{1 - 3 w}{12 (1 + w)} \ln \rpar{\frac{\rho_\text{th}}{\rho_\text{end}}} \, ,
\eeq

where $w$ is the average equation of state during reheating, which we will assume to be zero, and $\rho_\text{th}$ and $\rho_\text{end}$ are the energy densities of the universe at the point of thermalisation and the end of inflation, respectively. The two cases we study are $\rho_\text{th} = \rho_\text{end}$ (efficient reheating) and $\rho_\text{th} \ll \rho_\text{end}$ (inefficient reheating). For concreteness, we will choose for the latter case the ratio $\rho_\text{th} = e^{-60} \rho_\text{end}$ which corresponds to a shift in the observable window of 5 e-folds. We can see in the figure that while for both trajectories the tensor to scalar ratio is largely insensitive to the efficiency of reheating, the spectral index varies substantially, particularly in Trajectory B (fig \ref{fig:DBIB}) where particularly inefficient reheating could push the model outside of the 2$\sigma$ Planck contours shown on figure \ref{fig:nsr}. Trajectory A is much more stable under variations in the reheating process, however, with both cases considered making predictions comfortably in the $1\sigma$ contour. This is likely due to the fact that during the time when the observable scales leave the horizon during inflation for Trajectory A $\ga$ is roughly constant and so are the sound speeds. On the other hand, for Trajectory B $\ga$ has some appreciable evolution so that a shift in the observable window by 5 e-folds can significantly affect the scale dependence of the field perturbations.

In these results we neglect the possibility of further processing of the power spectrum due to post-inflationary isocurvature perturbations. The details of such a process would depend on the details of the decays during reheating, and is beyond the scope of the present work. This issue is discussed in section \ref{sec:postinflation}.

\section{Non-Gaussianities} \label{sec:NGcalculations}

To go beyond standard tests of inflation from the power spectrum, one can consider the three-point correlation function, and its related bispectrum, conventionally defined as,

\beq
\braket{\mathcal{R}(\mathbf{k}_1)\mathcal{R}(\mathbf{k}_2)\mathcal{R}(\mathbf{k}_3)}= (2 \pi)^3 \delta(\mathbf{k}_1 + \mathbf{k}_2 + \mathbf{k}_3) B(k_1, k_2, k_3) \, ,
\eeq

analogously to the well-known relation between the two point correlation and the usual power spectrum. Similarly, because of isotropy, the bispectrum depends only on the magnitudes of the three momenta, rather than their vector-valued forms. 
\newline
\newline
To compute leading order estimates of the non-Gaussianity predicted by disformally coupled inflation models, we use the In-In Formalism in the static approximation, such that quantities of interest (sound speeds, etc) are treated as nearly constant during horizon-crossing. While disformally coupled inflation is capable of giving rise to a wide range of phenomenology, including trajectories in which the sound speeds change rapidly up to and during horizon crossing, these trajectories typically do not lead to feasible power spectra as they generate large deviations from scale invariance which are heavily constrained by data. As such, we apply the static approximation in our calculations for simplicity with the understanding that the results are only accurate for reasonably slowly evolving trajectories. 
\newline
\newline
Analytically, however, this process would only be valid until roughly the time of horizon crossing, but we want to calculate non-Gaussianity present at the end of inflation, long after this. To deal with the fact that post-horizon-crossing the evolution of the two-field system of perturbations becomes more complex due to the presence of isocurvature perturbations, we shall parametrise the details of the late-time evolution of the system with the transfer function formalism \cite{Wands:2002bn,DiMarco:2002eb}. To do this, it is first convenient to define the adiabatic and entropic field perturbations as the linear combinations of the two fields which feed the curvature and isocurvature power spectrum, respectively. Essentially, this means we need to find a basis expansion, $e^I_n$,  of our fields $Q^I$ such that $Q^I = e^I_n Q^n$ with $n = (\sigma, s)$ for the adiabatic and entropy modes, respectively, which must be orthonormal in the sense that \cite{Langlois:2008qf},

\beq \label{eq:orthonormalcondition}
\bar{P}_{<IJ>}e^I_n e^J_m = \delta_{nm} \, ,
\eeq

where $\bar{P}_{<IJ>}$ denotes the matrix $P_{<IJ>}$ under the field redefinition $\chi \rightarrow \theta$. To explicitly find the elements of the matrix under this field redefinition, we directly expand $P_{<IJ>} Q^I Q^J$ ($I = \phi, \chi$) where $Q^\chi \rightarrow Q^\theta - \alpha Q^\phi$, 

\begin{align}
P_{<IJ>}Q^I Q^J & = P_{<IJ>}\rpar{\delta^I_\phi Q^{\phi} + \delta^I_\chi \spar{Q^\theta - \alpha Q^\phi}}\rpar{\delta^J_\phi Q^{\phi} + \delta^J_\chi \spar{Q^\theta - \alpha Q^\phi}} \, , \nonumber \\
& =  \rpar{P_\spp - 2 \alpha P_\spc + \alpha^2 P_\scc}Q^\phi Q^\phi + 2\rpar{P_\spc - \alpha P_\scc}Q^\phi Q^\theta + P_\scc Q^\theta Q^\theta \, .
\end{align}

From this it is clear that $\alpha = P_\spc / P_\scc = \ga^2 D \phidot \chidot /C$ is the choice of field redefinition which diagonalises the resulting matrix, which is precisely the result we found in eq. (\ref{eq:Qthetadef}). Inserting this into the above expression, we obtain the nonzero elements of the new matrix $\bar{P}_{<IJ>}$ (where now, $I,J = (\phi, \theta)$ rather than $(\phi, \chi)$), as,

\begin{align}
\bar{P}_\spp & = P_\spp - \frac{P_\spc^2}{P_\scc} = \ga\dbi- \ga^2 \frac{D}{C} p_\chi \\
\bar{P}_\stt & = P_\scc = \frac{C}{\ga} \, ,
\end{align}

where we have defined

\begin{align}
\ga\dbi=\left\{
                \begin{array}{ll}
                  1 \quad \text{in the canonical case (\ref{eq:CanonicalModel})} \\
                  \ga \quad \text{in the DBI case (\ref{eq:DBIModel})} \\
                \end{array}
              \right. \label{gammad}
\end{align}

in order to present the results for the two versions of the model in a unified fashion.
\newline
\newline
Similarly, the kinetic matrix (\ref{eq:KineticMatrix}) is diagonalised by this redefinition, with nonzero elements,

\begin{align}
\bar{K}\dpp & = \ga\dbi^3 +\ga^2 \frac{D}{C} \rho_\chi \\
\bar{K}\dtt &= \ga C \,.
\end{align}

The second-order action then contains the terms,

\beq
S_{(2)} \supset \int \bd t \bd^3 x a^3 \spar{\frac{1}{2} \bar{K}_{IJ} \dot{Q}^I \dot{Q}^J -\frac{1}{2}  \bar{P}_{<IJ>} h^{ij} \partial_i Q^I \partial_j Q^J} \,,
\eeq

from which we can see that the two sound speeds of the theory should be equal to the square roots of eigenvalues of $(\bar{K}^{-1})^{IK} \bar{P}_{<KJ>}$, which is easy to evaluate given the diagonal forms of these two matrices, and can hence be shown to agree with eqs. (\ref{eq:DBIthetasoundspeed}--\ref{eq:DBIphisoundspeed}) in the DBI case, and with our previous results in the canonical case, verifying this approach. Knowing the elements of $\bar{P}_{<IJ>}$, we can now solve the system in eq. (\ref{eq:orthonormalcondition}) for the orthonormal basis decomposition $e^I_n$.  Using the results of \cite{Longden:2016fgu}, we find 

\beq
e^I_\sigma = \frac{\varphidot^I}{\sqrt{\bar{P}_{<IJ>} \varphidot^I \varphidot^J}} \,,
\eeq

but the entropy vectors are a little less obvious. Still, solving the system (\ref{eq:orthonormalcondition}) using a reasonable ansatz we find,

\beq
e^\phi_s = - \sqrt{\frac{\bar{P}_\stt}{\bar{P}_\spp}} \frac{\thetadot}{\sqrt{\bar{P}_{<IJ>} \varphidot^I \varphidot^J}} \, , \quad e^\theta_s = \sqrt{\frac{\bar{P}_\spp}{\bar{P}_\stt}} \frac{\phidot}{\sqrt{\bar{P}_{<IJ>} \varphidot^I \varphidot^J}} \, .
\eeq

From this we find we can write the adiabatic and entropy perturbations as,

\begin{align}
Q^\sigma &= \frac{\bar{P}_{<IJ>} \varphidot^I Q^J}{\sqrt{\bar{P}_{<IJ>} \varphidot^I \varphidot^J}} \, ,  \label{eq:Qsigmadef} \\
Q^s & = \sqrt{\frac{|\bar{P}|}{\bar{P}_{<IJ>} \varphidot^I \varphidot^J}} \rpar{\thetadot Q^\phi - \phidot Q^\theta}  \label{eq:Qsdef} \, ,
\end{align}

where by $|\bar{P}|$ we mean the determinant of $\bar{P}_{<IJ>}$, or more explicitly $\bar{P}_\spp \bar{P}_\stt = P_\spp P_\scc - P_\spc^2$. These quantities then independently source the adiabatic curvature perturbation and the entropy perturbation, respectively, for which we take the definitions

\begin{align}
\mathcal{R} & = \frac{H}{\sqrt{\bar{P}_{<IJ>} \varphidot^I \varphidot^J}} Q^\sigma \, , \\
\mathcal{S} & = \mathcal{S}_0 \frac{H}{\sqrt{\bar{P}_{<IJ>} \varphidot^I \varphidot^J}} Q^s \, ,
\end{align}

where we have included the factor $\mathcal{S}_0$ in our definition of the entropy perturbation so that its value can be chosen to ensure that the power spectra of curvature and entropy modes at horizon crossing are equal, as this simplifies the following calculations without loss of generality. To realise this, it should take the value

\begin{align}
\mathcal{S}_0
 = \frac{Q_\sigma^*}{Q_s^*}  
 = \frac{1}{|\bar{P}|^{1/2}}\frac{\bar{P}\dpp \phidot Q_\phi^* + \bar{P}\dtt \thetadot Q_\theta^*}{\thetadot Q_\phi^*  - \phidot Q_\theta^* } \, .
\end{align}

We can estimate the value of $Q_\phi$ and $Q_\theta$ at horizon crossing by using the $\eta \rightarrow 0 $ limit of the early-time solutions of the equation of motion that are asymptotic to the Bunch-Davies vacuum state, which obey the standard result,

\beq
Q_I = \frac{H}{\sqrt{2 k^3 c_s^I}} (1 - i k c_s^I \eta) e^{i k c_s^I \eta} \,.
\eeq

Making use of this form of the $Q_\phi$ and $Q_\theta$, we obtain

\beq
\mathcal{S}_0 = \frac{1}{|\bar{P}|^{1/2}} \frac{\bar{P}\dpp \phidot \sqrt{c_s^{(\theta)}} + \bar{P}\dtt \thetadot \sqrt{c_s^{(\phi)}}}{\thetadot \sqrt{c_s^{(\theta)}}  - \phidot \sqrt{c_s^{(\phi)}}} \, .
\eeq

Note that while we have found the adiabatic and entropy perturbations for their convenience in application of the transfer function formalism, we emphasise that they are not necessarily the most useful fields in which one might like to think about the system. Indeed, if one is concerned with brane world scenarios which give rise to disformally coupled inflation, the fields $\phi$ and $\chi$ hold more physical meaning. Alternatively, if one is concerned with the independent fields which each propagate with a well defined sound speed, then that is the $(\phi,\theta)$ basis, as it is in terms of these fields that the sound-speed matrix is diagonal, as argued above. This ($\phi$, $\theta$) basis will also prove the most convenient for performing calculations of power spectra, as when it comes to quantisation of the perturbations, these are the modes we can canonically quantise each with a single speed of sound. It will hence be helpful to express the final curvature perturbation in this basis. This results in, 

\begin{align}
\mathcal{R}_{\text{end}}
& = \mathcal{R}^*+ \TRS \mathcal{S}^*  \nonumber \\
& = \frac{H}{\sqrt{\bar{P}_{<IJ>} \varphidot^I \varphidot^J}} \rpar{Q^\sigma_* + \mathcal{S}_0 \mathcal{T}_{\mathcal{RS}}  Q^s_*  } \nonumber \\
& =  \frac{H}{\bar{P}_{<IJ>} \varphidot^I \varphidot^J} \rpar{ \spar{\bar{P}\dpp \phidot + A \thetadot} Q^\phi_* +  \spar{\bar{P}\dtt \thetadot -  A \phidot}  Q^\theta_* } \nonumber \\
& = F_\phi Q^\phi_* + F_\theta Q^\theta_* \, . \label{eq:RfromQs}
\end{align}

where, recalling the definition of $\ga\dbi$ in (\ref{gammad}), we have

\begin{align} \label{eq:Fcoefficients1}
F_\phi & = \frac{H}{P_{<IJ>} \phidot^I \phidot^J} \spar{\rpar{\ga\dbi - \ga^2 \frac{D}{C}p_\chi} \phidot + A \thetadot}\,,  \\
F_\theta & =\frac{H}{P_{<IJ>} \phidot^I \phidot^J}  \spar{\frac{C}{\ga} \thetadot -  A \phidot } \,, \label{eq:Fcoefficients2}
\end{align}

and

\begin{align}
A 
& = \mathcal{S}_0 \TRS |\bar{P}|^{1/2}  \\
& = \TRS \frac{\bar{P}\dpp \phidot \sqrt{c_s^{(\theta)}} + \bar{P}\dtt \thetadot \sqrt{c_s^{(\phi)}}}{\thetadot \sqrt{c_s^{(\theta)}}  - \phidot \sqrt{c_s^{(\phi)}}}  \\
& = \TRS \frac{\ga \rpar{\ga\dbi - \ga^2 \frac{D}{C} p_\chi}\phidot \sqrt{c_s^{(\theta)}} + C \thetadot \sqrt{c_s^{(\phi)}}}{\ga \thetadot \sqrt{c_s^{(\theta)}}  - \ga \phidot \sqrt{c_s^{(\phi)}}} \,.
\end{align}

In the above expressions, it is to be understood that all background quantities are evaluated at horizon crossing even though we have omitted explicitly marking them as such for the sake of readability. Only the transfer function $\TRS$ depends on the details of inflation after a given mode has left the horizon. 
\newline
\newline
We are now in the position to use the In-In formalism to compute the three-point function up to the time of horizon-crossing when our analytical approximations are still accurate, and then use the language of the transfer function as applied in the above discussion to relate our results to the three-point statistics of the curvature perturbation at the end of inflation. Using for example eq. (\ref{eq:RfromQs}), we can convert three-point functions of the field perturbations (which can be calculated from the third order action) into three-point functions of the curvature perturbation, via the relation

\begin{align}
\tpcfrrr & = F_\phi^3 \tpcfppp \nonumber \\ & + F_\phi^2 F_\theta (\tpcfppc \pperms) \nonumber \\ & + F_\phi F_\theta^2 (\tpcfpcc \pperms)  \nonumber \\ & + F_\theta^3 \tpcfccc \\ \nonumber \\
& =  F_\phi^3 \tpcfppp + F_\phi^2 F_\theta \ppcperms \nonumber \\
& + F_\phi F_\theta^2 \pccperms + F_\theta^3 \tpcfccc \, , \label{eq:RfromQs3}
\end{align}

where `$ \pperms$' indicates the inclusion of similar terms with all distinct permutations of the momenta $\mathbf{k}_n$, and for convenience we have used the shorthands:

\begin{align}
\ppcperms & = (\tpcfppc \pperms) \, ,\\
\pccperms & = (\tpcfpcc \pperms) \, .
\end{align}

However, as the $F_\phi$ and $F_\theta$ coefficients in eq. (\ref{eq:RfromQs3}) contain factors of the transfer function $\TRS$, we need a way of computing this. Fortunately, this can be obtained from our numerical simulations of the perturbed equations of motion carried out in section \ref{sec:DBIspec} and our previous paper \cite{vandeBruck:2015tna}, by comparing the final value of $\mathcal{P}_\mathcal{R}$ to its value at horizon-crossing.
\newline
\newline
While the full third-order action is rather complicated, it has fortunately been studied previously \cite{Langlois:2008qf,Arroja:2008yy} which terms make the leading order contribution to non-Gaussianities. We hence work with the subset of the third-order action which contains the relevant vertices for this leading order calculation, that is,

\beq \label{eq:toactionx}
\mathcal{L}_3 \supset a^3 g_{IJK} \dot{Q}^I \dot{Q}^J \dot{Q}^K + a h_{IJK} \dot{Q}^I \partial_i Q^J \partial^i Q^K \,,
\eeq

with

\begin{align}
g_{IJK} & = \frac{1}{2} P_{<JK><AI>} \phidot^A + \frac{1}{6} P_{<AI><BJ><CK>} \phidot^A \phidot^B \phidot^C \, , \label{eq:gcoefficients} \\
h_{IJK} & = -\frac{1}{2} P_{<JK><AI>} \phidot^A  \, . \label{eq:hcoefficients}
\end{align}

We want to rewrite this in terms of the ($\phi, \theta$) basis before we proceed to calculate three-point functions, however, as it is these fields which have well-defined sound speeds and are most suitable for quantisation. To this effect, we perform the field redefinition (\ref{eq:Qthetadef}) to obtain the modified third order action

\beq \label{eq:toactionx2}
\mathcal{L}_3 \supset a^3 g_{I'J'K'} \dot{Q}^{I'} \dot{Q}^{J'} \dot{Q}^{K'} + a h_{I'J'K'} \dot{Q}^{I'} \partial_i Q^{J'} \partial^i Q^{K'} \,,
\eeq

where primed indices take values from $(\phi, \theta)$ instead of $(\phi, \chi)$. The new coefficients will take the form

\beq
f_{I'J'K'} = e^{I}_{I'} e^{J}_{J'} e^{K}_{K'} f_{IJK} \, , \quad  (f = g, h) \, , \label{eq:primedcoeffs}
\eeq

where $e^\phi_\phi = e^\chi_\chi = 1$, $e^\phi_\theta = 0$ and $e^\chi_\phi = - \ga^2 D \phidot \chidot / C$ from eq. (\ref{eq:Qthetadef}).
\newline
\newline
There are hence 16 important terms in the third order action, corresponding to the 8 permutations of $IJK$, each with two vertices. Note that as $P_{\scc<A\chi>} = P_{<A\chi><B\chi><C\chi>} = 0$ for any values of $A$, $B$ and $C$ due to the Lagrangian being at most second order in $X\pc$ and first order in $X\cc$, both $g_{\chi\chi\chi}$ and $h_{\chi\chi\chi}$ are identically zero, and no non-Gaussianity is generated purely due to $\chi$ at leading order. From eq. (\ref{eq:primedcoeffs}) we can see that $g_{\theta \theta \theta} = g_{\chi \chi \chi}$ and $h_{\theta \theta \theta} = h_{\chi \chi \chi}$, so it is clear that the term $\tpcfccc$ in eq. (\ref{eq:RfromQs3}) will be zero. The non-zero terms are those involving just $\phi$, due to the Lagrangian's irregular dependence on its kinetic term through $\gamma$, and those terms involving $\phi$ and $\theta$ due to their interactions, which are also highly non-standard. We will hence have non-zero $\tpcfppp$, $\tpcfppc$ and $\tpcfpcc$ at this order.

\subsection{Computing the leading-order bispectrum}

To compute three-point functions we use the standard leading order result \cite{Maldacena:2002vr}

\beq
\braket{Q^I(t,\mathbf{k}_1) Q^J(t,\mathbf{k}_2) Q^K(t,\mathbf{k}_3)} = -i \int_{t_0}^t\Braket{\spar{Q^I(t,\mathbf{k}_1) Q^J(t,\mathbf{k}_2) Q^K(t,\mathbf{k}_3), H_\text{int}(t')}} \bd t' \, ,
\eeq

where $t_0$ implies some early time when fluctuations are deep within the horizon. In practise we use conformal time $\eta$ (where $a \bd \eta = \bd t$) with which it is a reasonable approximation to perform the above integral between the limits $[-\infty,0]$ as is usually done.  $H_\text{int}$ is the interaction Hamiltonian, which at this level of approximation can be identified as $-\mathcal{L}_3$, the third order perturbed Lagrangian, or rather for our purposes, the subset of it which we identified in eq. (\ref{eq:toactionx}) as providing the main contribution to the non-Gaussianity.
\newline
\newline
We will specialise our results to the equilateral configuration of non-Gaussianity, to give a concrete and simple example which can be numerically evaluated. We choose the equilateral configuration for its simplicity in that all three momenta are equal and one does not need to account for effects such as one mode being far outside the horizon while the others are crossing it, as seen for example in \cite{Seery:2005wm}. We emphasise that this choice is not meant to imply that it is the most significant or interesting configuration for our model. 


\subsubsection{Kinetic vertices}

To be explicit, we write out the kinetic terms in the reduced third-order action (\ref{eq:toactionx2}) as (noting that $g_{\theta \theta \theta} = g_{\chi \chi \chi} = 0$)

\beq
a^3 g_{\phi\phi\phi} (\dot{Q}^\phi)^3 + a^3 (g_{\phi\phi\theta}+g_{\phi\theta\phi}+g_{\theta\phi\phi}) (\dot{Q}^\phi)^2 \dot{Q}^\theta+ a^3 (g_{\phi\theta\theta}+g_{\theta\phi\theta}+g_{\theta\theta\phi}) \dot{Q}^\phi (\dot{Q}^\theta)^2 \, .
\eeq

For the vertex $a^3 g_{\phi\phi\phi} (\dot{Q}^\phi)^3$, we obtain a contribution to the $\tpcfppp$ correlation function of the form,

\beq
\tpcfppp \supset (2 \pi)^3 \delta(\mathbf{k}_1 + \mathbf{k}_2 + \mathbf{k}_3) 3 g_{\phi\phi\phi} H^5 \frac{1}{\prod k_i^3} \frac{k_1^2 k_2^2 k_3^2}{K^3}\,.
\eeq

Similarly for the interaction $a^3 (g_{\phi\phi\theta}+g_{\phi\theta\phi}+g_{\theta\phi\phi}) (\dot{Q}^\phi)^2 \dot{Q}^\theta$, we find

\beq
\ppcperms \supset (2 \pi)^3 \delta\rpar{\sum \mathbf{k}}(g_{\phi\phi\theta}+g_{\phi\theta\phi}+g_{\theta\phi\phi}) H^5 \frac{k_1^2 k_2^2 k_3^2 (c_s^\phi)^2 c_s^\theta }{\prod k_i^3}  \Bigg[\frac{1}{(\kappa^{\phi\phi\theta})^3}+\frac{1}{(\kappa^{\phi\theta\phi})^3} +\frac{1}{(\kappa^{\theta\phi\phi})^3}\Bigg] \, ,
\eeq

and for the vertex $a^3 (g_{\phi\theta\theta}+g_{\theta\phi\theta}+g_{\theta\theta\phi}) \dot{Q}^\phi (\dot{Q}^\theta)^2$,

\beq
\pccperms \supset (2 \pi)^3 \delta\rpar{\sum \mathbf{k}}(g_{\phi\theta\theta}+g_{\theta\phi\theta}+g_{\theta\theta\phi}) H^5 \frac{ k_1^2 k_2^2 k_3^2 (c_s^\theta)^2 c_s^\phi }{\prod k_i^3} \Bigg[\frac{1}{(\kappa^{\phi\theta\theta})^3}+\frac{1}{(\kappa^{\theta\phi\theta})^3} +\frac{1}{(\kappa^{\theta\theta\phi})^3}\Bigg] \, .
\eeq

We have used the shorthands $K = k_1 + k_2 + k_3$ and $\kappa^{IJK} = c_s^I k_1 + c_s^J k_2 + c_s^K k_3$ in these expressions. In the equilateral configuration $k_1=k_2=k_3 = k$, so $\kappa^{IJK} = k \rpar{c_s^I + c_s^J + c_s^K}$ and $K = 3k$, and the above expressions reduce to:

\beq
\tpcfppp \supset (2 \pi)^3 \delta\rpar{\sum \mathbf{k}} \frac{1}{9} g_{\phi\phi\phi} \frac{H^5}{k^6} \, ,
\eeq

\beq
\ppcperms \supset (2 \pi)^3 \delta\rpar{\sum \mathbf{k}} 3 (g_{\phi\phi\theta}+g_{\phi\theta\phi}+g_{\theta\phi\phi}) \frac{H^5}{k^6}  \rpar{\frac{(c_s^\phi)^2 c_s^\theta }{(2 c_s^\phi + c_s^\theta)^3}} \, ,
\eeq

and 

\beq
\pccperms \supset (2 \pi)^3 \delta\rpar{\sum \mathbf{k}} 3 (g_{\phi\theta\theta}+g_{\theta\phi\theta}+g_{\theta\theta\phi}) \frac{H^5}{k^6}  \rpar{\frac{(c_s^\theta)^2 c_s^\phi }{(2 c_s^\theta + c_s^\phi)^3}} \, .
\eeq

\subsubsection{Gradient vertices}

In Fourier space, spatial derivatives correspond to factors of $i \mathbf{k}$. In particular since the third order action contains contractions of spatial derivatives two fields we will pick up a factor such as $- \mathbf{k_1} \cdot \mathbf{k_2}$ from these terms. Noting the isotropy condition that $\mathbf{k}_1 + \mathbf{k}_2 + \mathbf{k}_3 = 0$, writing $k_1^2 = \mathbf{k_1} \cdot \mathbf{k_1} = (\mathbf{k_2}+\mathbf{k_3}) \cdot (\mathbf{k_2}+\mathbf{k_3}) = k_2^2 + k_3^2 + 2 \mathbf{k_2}\cdot\mathbf{k_3}$, we identify that $\mathbf{k_2}\cdot\mathbf{k_3} = (k_1^2 - k_2^2 - k_3^2)/2$, which, along with equivalent expressions for the other permutations of the three $k$s, will be later useful in evaluating these expressions. In this section we will introduce for convenience the shorthand

%

\beq
F^{I_1 I_2 I_3}_{n} = 2 (\kappa^{I_1 I_2 I_3})^2 - \bar{k}_n \kappa^{I_1 I_2 I_3} + \delta_n^{ml} \bar{k}_m \bar{k}_l \, ,
\eeq

where $n \in [1,2,3]$ and $\delta_n^{ml} = 1$ when $n \neq m \neq l$ and $0$ otherwise. We also define here $\bar{k}_n = c_s^{I_n} k_n$. When all three fields are the same we use the shorthand $F_n = F^{III}_n/(c_s^I)^2 = 2 K^2 - k_n K + \delta_n^{ml} k_m k_l$.
\newline
\newline
Our action, explicitly written, contains the terms (noting that $h_{\theta \theta \theta} = h_{\chi \chi \chi} = 0$),

\begin{align}
a h_{\phi\phi\phi} \dot{Q}^\phi (\partial Q^\phi)^2 & + a (h_{\phi\phi\theta}+h_{\phi\theta\phi})\dot{Q}^\phi \partial_i Q^\phi \partial^i Q^\theta + a h_{\theta\phi\phi} \dot{Q}^\theta (\partial Q^\phi)^2 \\
& + a (h_{\theta\phi\theta}+h_{\theta\theta\phi})\dot{Q}^\theta \partial_i Q^\phi \partial^i Q^\theta + a h_{\phi\theta\theta} \dot{Q}^\phi (\partial Q^\theta)^2 \, .
\end{align}

For the vertex $a h_{\phi\phi\phi} \dot{Q}^\phi (\partial Q^\phi)^2$ there is a contribution of the form,

\beq
\tpcfppp \supset (2 \pi)^3 \delta\rpar{\sum \mathbf{k}}  \frac{h_{\phi\phi\phi} H^5}{2 (c_s^\phi)^2} \frac{1}{\prod k_i^3} \frac{1}{K^3}
 \Big[ 
 k_1^2 (\mathbf{k_2}\cdot\mathbf{k_3})  F_1 + k_2^2 (\mathbf{k_3}\cdot\mathbf{k_1})  F_2  + k_3^2 (\mathbf{k_1}\cdot\mathbf{k_2}) F_3
\Big] \, ,
\eeq

while for the vertex $a h_{\phi\theta\theta} \dot{Q}^\phi (\partial Q^\theta)^2$, we have,

\beq
\pccperms \supset (2 \pi)^3 \delta\rpar{\sum \mathbf{k}} \frac{h_{\phi\theta\theta} H^5}{2} \frac{1}{\prod k_i^3} \frac{c_s^\phi}{(c_s^\theta)^2}  \Bigg[ k_1^2 (\mathbf{k_2}\cdot\mathbf{k_3})\frac{F^{\phi\theta\theta}_{1}}{(\kappa^{\phi\theta\theta})^3} +  k_2^2 (\mathbf{k_3}\cdot\mathbf{k_1}) \frac{F^{\theta\phi\theta}_{2}}{(\kappa^{\theta\phi\theta})^3}+ k_3^2 (\mathbf{k_1}\cdot\mathbf{k_2}) \frac{F^{\theta\theta\phi}_{3}}{(\kappa^{\theta\theta\phi})^3}  \Bigg] \, .
\eeq

Similarly, for $a h_{\theta\phi\phi} \dot{Q}^\theta (\partial Q^\phi)^2$, 

\beq
\ppcperms \supset (2 \pi)^3 \delta\rpar{\sum \mathbf{k}} \frac{h_{\theta\phi\phi} H^5}{2} \frac{1}{\prod k_i^3} \frac{c_s^\theta}{(c_s^\phi)^2}  \Bigg[ k_1^2 (\mathbf{k_2}\cdot\mathbf{k_3})\frac{F^{\theta\phi\phi}_{1}}{(\kappa^{\theta\phi\phi})^3} +  k_2^2 (\mathbf{k_3}\cdot\mathbf{k_1}) \frac{F^{\phi\theta\phi}_{2}}{(\kappa^{\phi\theta\phi})^3}+ k_3^2 (\mathbf{k_1}\cdot\mathbf{k_2}) \frac{F^{\phi\phi\theta}_{3}}{(\kappa^{\phi\phi\theta})^3}  \Bigg] \, .
\eeq

For $ a (h_{\phi\phi\theta}+h_{\phi\theta\phi})\dot{Q}^\phi \partial_i Q^\phi \partial^i Q^\theta$, we obtain

\begin{align}
\ppcperms \supset (2 \pi)^3 \delta\rpar{\sum \mathbf{k}} \frac{(h_{\phi\phi\theta}+h_{\phi\theta\phi}) H^5}{4} \frac{1}{\prod k_i^3} \frac{1}{c_s^\theta} \Bigg[ &
\frac{k_2^2 (\mathbf{k_3}\cdot\mathbf{k_1}) F^{\theta\phi\phi}_2 + k_3^2 (\mathbf{k_1}\cdot\mathbf{k_2}) F^{\theta\phi\phi}_3}{(\kappa^{\theta\phi\phi})^3} \nonumber \\
& \frac{k_3^2 (\mathbf{k_1}\cdot\mathbf{k_2}) F^{\phi\theta\phi}_3 +  k_1^2 (\mathbf{k_2}\cdot\mathbf{k_3}) F^{\phi\theta\phi}_1}{(\kappa^{\phi\theta\phi})^3} \nonumber \\
& \frac{ k_1^2 (\mathbf{k_2}\cdot\mathbf{k_3}) F^{\phi\phi\theta}_1 + k_2^2 (\mathbf{k_3}\cdot\mathbf{k_1}) F^{\phi\phi\theta}_2}{(\kappa^{\phi\phi\theta})^3}
\Bigg] \, .
\end{align}

and finally the vertex $a (h_{\theta\phi\theta}+h_{\theta\theta\phi})\dot{Q}^\theta \partial_i Q^\phi \partial^i Q^\theta $ leads to,

\begin{align}
\pccperms \supset (2 \pi)^3 \delta\rpar{\sum \mathbf{k}} \frac{(h_{\theta\phi\theta}+h_{\theta\theta\phi}) H^5}{4} \frac{1}{\prod k_i^3} \frac{1}{c_s^\phi} \Bigg[ &
\frac{k_2^2 (\mathbf{k_3}\cdot\mathbf{k_1}) F^{\phi\theta\theta}_2 + k_3^2 (\mathbf{k_1}\cdot\mathbf{k_2}) F^{\phi\theta\theta}_3}{(\kappa^{\phi\theta\theta})^3} \nonumber \\
& \frac{k_3^2 (\mathbf{k_1}\cdot\mathbf{k_2}) F^{\theta\phi\theta}_3 +  k_1^2 (\mathbf{k_2}\cdot\mathbf{k_3}) F^{\theta\phi\theta}_1}{(\kappa^{\theta\phi\theta})^3} \nonumber \\
& \frac{ k_1^2 (\mathbf{k_2}\cdot\mathbf{k_3}) F^{\theta\theta\phi}_1 + k_2^2 (\mathbf{k_3}\cdot\mathbf{k_1}) F^{\theta\theta\phi}_2}{(\kappa^{\theta\theta\phi})^3}
\Bigg] \, .
\end{align}

Writing all of these results in the equilateral configuration, where 

\beq
F^{IJJ}_{n\text{(eq)}} = F^{JIJ}_{n\text{(eq)}} = F^{JJI}_{n\text{(eq)}} = \spar{(c_s^I)^2 + 10 (c_s^J)^2 + 6 c_s^I c_s^J}k^2 \, ,
\eeq

\beq
F^{IIJ}_{n\text{(eq)}} = F^{IJI}_{n\text{(eq)}} = F^{JII}_{n\text{(eq)}} = \spar{6 (c_s^I)^2 + 2 (c_s^J)^2 + 9 c_s^I c_s^J}k^2 \, ,
\eeq

and $F_n = 17 k^2$ then gives us,

\beq
\tpcfppp \supset -(2 \pi)^3 \delta\rpar{\sum \mathbf{k}} \frac{17}{36} h_{\phi\phi\phi} \frac{H^5}{k^6} \frac{1}{(c_s^\phi)^2} \,,
\eeq

\beq
\pccperms \supset -(2 \pi)^3 \delta\rpar{\sum \mathbf{k}} \frac{3}{4} h_{\phi\theta\theta} \frac{H^5}{k^6} \rpar{\frac{c_s^\phi\spar{(c_s^\phi)^2 + 10 (c_s^\theta)^2 + 6 c_s^\phi c_s^\theta}}{(c_s^\theta)^2(2 c_s^\theta + c_s^\phi)^3}} \,,
\eeq

\beq
\ppcperms \supset -(2 \pi)^3 \delta\rpar{\sum \mathbf{k}} \frac{3}{4} h_{\theta\phi\phi} \frac{H^5}{k^6} \rpar{\frac{c_s^\theta\spar{(c_s^\theta)^2 + 10 (c_s^\phi)^2 + 6 c_s^\phi c_s^\theta}}{(c_s^\phi)^2(2 c_s^\phi + c_s^\theta)^3}} \,,
\eeq

\beq
\ppcperms \supset -(2 \pi)^3 \delta\rpar{\sum \mathbf{k}} \frac{3}{4} (h_{\phi\phi\theta}+h_{\phi\theta\phi}) \frac{H^5}{k^6}\rpar{\frac{6 (c_s^\phi)^2 + 2 (c_s^\theta)^2 + 9 c_s^\phi c_s^\theta}{c_s^\theta (2 c_s^\phi + c_s^\theta)^3}} \,,
\eeq

\beq
\pccperms \supset -(2 \pi)^3 \delta\rpar{\sum \mathbf{k}} \frac{3}{4} (h_{\theta\theta\phi}+h_{\theta\phi\theta}) \frac{H^5}{k^6}\rpar{\frac{6 (c_s^\theta)^2 + 2 (c_s^\phi)^2 + 9 c_s^\phi c_s^\theta}{c_s^\phi (2 c_s^\theta + c_s^\phi)^3}} \,.
\eeq

\subsubsection{Total non-Gaussianity}

Summing all the previously calculated contributions, in the equilateral configuration, we hence have leading order three-point functions of the fields in the form,

\beq \label{eq:pppeqresult}
\tpcfppp = (2 \pi)^3 \delta\rpar{\sum \mathbf{k}}\frac{H^5}{k^6} \rpar{ \frac{1}{36} \spar{4 g_{\phi\phi\phi}- \frac{17 h_{\phi\phi\phi}}{(c_s^\phi)^2} } } \, ,
\eeq

\beq \label{eq:ppceqresult}
\ppcperms = (2 \pi)^3 \delta\rpar{\sum \mathbf{k}}\frac{H^5}{k^6} \rpar{\frac{3}{4} \frac{s}{ (2 + s)^3} \Bigg[4 g_1 - h_{\theta\phi\phi} \frac{s^2 + 6s +10}{(c_s^\phi)^2} - h_1 \frac{2s^2 + 9s + 6}{(c_s^\theta)^2} \Bigg] }\, ,
\eeq

\beq \label{eq:pcceqresult}
\pccperms = (2 \pi)^3 \delta\rpar{\sum \mathbf{k}}\frac{H^5}{k^6} \rpar{\frac{3}{4} \frac{\bar{s}}{ (2 + \bar{s})^3} \Bigg[4 g_2 - h_{\phi\theta\theta} \frac{\bar{s}^2 + 6\bar{s} +10}{(c_s^\theta)^2} - h_2 \frac{2\bar{s}^2 + 9\bar{s}+ 6}{(c_s^\phi)^2} \Bigg]} \, ,
\eeq

where 

\begin{align*}
&g_1 = g_{\phi\phi\theta}+g_{\phi\theta\phi}+g_{\theta\phi\phi} \, , & \quad & g_2 = g_{\phi\theta\theta}+g_{\theta\phi\theta}+g_{\theta\theta\phi} \,  , \\
&h_1  = h_{\phi\phi\theta}+h_{\phi\theta\phi} \, , & \quad & h_2  = h_{\phi\theta\theta}+h_{\theta\phi\theta} \, , \\
&s  = \frac{c_s^\theta}{c_s^\phi} \, , & \quad  & \bar{s} = \frac{1}{s} \, .
\end{align*}

and the $g$ and $h$ coefficients are given in eq. (\ref{eq:primedcoeffs}). Using this, we can then evaluate the expression for $\tpcfrrr$, given in eq. (\ref{eq:RfromQs3}) using our numerical simulations. It is conventional, and more useful, however to convert the numerical value of $\tpcfrrr$ into an $f_{NL}$ value, such as via the relation \cite{Langlois:2008qf},

\beq
\tpcfrrr = - (2 \pi)^7 \delta\rpar{\sum \mathbf{k}} \frac{\sum k_i^3}{\prod k_i^3} \rpar{\frac{3}{10} f_{NL} \mathcal{P}_\mathcal{R}^2} \, .
\eeq

As in the above calculations of the power spectrum, we calculate $f_{NL}$ at the end of inflation, neglecting the possibility (depending on the details of reheating) of effects due to the post-inflationary persistence of isocurvature perturbations. This is discussed in section \ref{sec:postinflation}. In table \ref{table:1} we present the calculated $f_{NL}$ values from the end of inflation along with corresponding power spectrum properties for various trajectories studied for the canonical and DBI models.
\newline
\newline
\begin{table}[]
\centering
\caption{The calculated equilateral non-Gaussianity( $f_{NL}$) in various models studied, as well as the amplitude ($A_s$), tilt ($n_s$), tensor-to-scalar-ratio ($r$), running ($\alpha_s$) and running of the running ($\beta_s$) for inflationary trajectories studied above and in previous work.}
\label{table:1}
\begin{tabular}{|r|c|c|c|c|c|c|}
\hline
\multicolumn{1}{|c|}{\textbf{Trajectory}} & \textbf{$10^9 A_s $} & \textbf{$n_s$} & \textbf{$10^4 \alpha_s$} & \textbf{$10^4 \beta_s$} & \textbf{$10^3 r$} & \textbf{$f_{NL}$} \\ \hline
Canonical A\cite{vandeBruck:2015tna}                               & 2.12                 & 0.961          & -5.3            & 1.8           & 17        & -29.5                \\ \hline
Canonical B\cite{vandeBruck:2015tna}                               & 2.15                 & 0.968          & 7.1            & -0.21        & 31       & -0.33             \\ \hline
Canonical C\cite{vandeBruck:2015tna}                              & 2.15                 & 0.967          & 12            & -11            & $1.2 \times 10^{-6}$ & $-2.4 \times 10^6$             \\ \hline
DBI A (Fig. \ref{fig:DBIA})                                                    & 2.14                 & 0.965          & 2.4            & 2.1            & 7.2      & -0.59           \\ \hline
DBI B    (Fig. \ref{fig:DBIB})                                                  & 2.14                 & 0.973          & -59             & -0.64         & 2.0       & 0.88            \\ \hline
\end{tabular}
\end{table}
Our results show that it is feasible for specific parameters of disformally coupled inflation to simultaneously predict a realistic power spectrum with either small or large non-Gaussianity, though we most often find it to be small and negative. The major exception to this is canonical trajectory C, which produces excessive non-Gaussianity. While it's power spectrum is in good agreement with the experimental data, it's bispectrum rules it out as a feasible inflationary model. Many things affect the amount of non-Gaussianity, including the sound speeds of the two fields and their ratio, the magnitude of the transfer function $\TRS$, and the size of the disformal effects via their influence on factors such as $F_\phi$ and $F_\theta$. In general, we see that a large deviation from $c_s = 1$ will amplify the non-gaussianity as usual, due to the factors of $c_s^{-2}$ in eqs. (\ref{eq:pppeqresult}--\ref{eq:pcceqresult}). This is one factor in explaining why canonical trajectory C produces such large non-Gaussianity - the parameter $\gamma$ is much larger than in all the other studied trajectories and hence the sound speeds of the two fields are close to zero. The ratio of sound-speeds $s$ and its reciprocal $\bar{s}$ influence the $\ppcperms$ and $\pccperms$ vertices, respectively. When $c_s^\phi = c_s^\theta = c_s$, and so $s = \bar{s} =  1$, the expressions (\ref{eq:ppceqresult}--\ref{eq:pcceqresult}) take on the forms

\beq \label{eq:ppceqresults1}
\ppcperms = (2 \pi)^3 \delta\rpar{\sum \mathbf{k}}\frac{H^5}{k^6} \rpar{\frac{1}{36} \Bigg[4 g_1 -  \frac{17 (h_{\theta\phi\phi} + h_{\phi\theta\phi} + h_{\phi\phi\theta})}{c_s^2}\Bigg] }\, ,
\eeq

and

\beq \label{eq:pcceqresults1}
\pccperms = (2 \pi)^3 \delta\rpar{\sum \mathbf{k}}\frac{H^5}{k^6} \rpar{\frac{1}{36} \Bigg[4 g_2 -  \frac{17 (h_{\theta\theta\phi} + h_{\theta\phi\theta} + h_{\phi\theta\theta})}{c_s^2} \Bigg]} \, ,
\eeq

which look much more similar to the $\tpcfppp$ result in eq. (\ref{eq:pppeqresult}). However, if we consider the limit $c_s^\theta > c_s^\phi$, or $s \gg 1$ the expressions reduce to (keeping only the term proportional to $(c_s^\phi)^{-2}$ as in this limit it should be much larger than the other),

\beq \label{eq:ppceqresultsbig}
\ppcperms = (2 \pi)^3 \delta\rpar{\sum \mathbf{k}}\frac{H^5}{k^6} \rpar{\frac{1}{36} \Bigg[108 g_1 \bar{s}^2 -  \frac{27 h_{\theta\phi\phi}}{(c_s^\phi)^2} \Bigg] }\, ,
\eeq

\beq \label{eq:pcceqresultsbig}
\pccperms = (2 \pi)^3 \delta\rpar{\sum \mathbf{k}}\frac{H^5}{k^6} \rpar{\frac{1}{36} \Bigg[108 g_2 s^2 -  \frac{54 h_2}{(c_s^\phi)^2} \Bigg]} \, .
\eeq

We see here in the $\pccperms$ result that now the $g_2$ term is amplified by a factor of $s^2$ and may provide a very large contribution in this limit. Similar expressions can be found in the opposite limit of $\bar{s} \gg 1$. This is not, however, achieved in any of the discussed trajectories, with all of them producing ratios of sound speeds close to unity at horizon crossing. We find a condition for deviation from $s \approx 1$ in the form

\beq \label{eq:soundspeeddiff}
(c_s^\theta)^2 - (c_s^\phi)^2  \approx \frac{1}{\ga^2} + w_\chi \, , 
\eeq

which is valid when disformal effects are large, in the sense that $D \rho_\chi> \ga C$ (for the DBI case) or $\ga^2 D \rho_\chi > C$  (for the canonical case). The equation of state for $\chi$ is given by

\beq
w_\chi = \frac{1}{\ga^2} \frac{\ga^2 X\cc - CV}{\ga^2 X\cc + CV} \, .
\eeq

Notice that when $X\cc = 0$ and $C V \neq 0$, this is equal to $-1/\ga^2$ and thus we would expect equal sound speeds according to eq. (\ref{eq:soundspeeddiff}). This implies that for non-trivial realisations of the model where the disformal coupling plays an important role in the dynamics, deviation from equal sound speeds occurs when the $\chi$ field is not potential-dominated. We may hence expect that variations of this model in which the $\chi$ field is allowed to fast-roll during inflation, non-Gaussianities may be amplified due to e.g. the term proportional to $s^2$ in eq. (\ref{eq:pcceqresultsbig}).
\newline
\newline
With regards to the large-scale evolution of the perturbations, parametrised by the transfer function $\TRS$, we can see that $f_{NL}$ goes roughly as $\tpcfrrr/\mathcal{P}_\mathcal{R}^2$. The effect of the transfer function on $\tpcfrrr$ is encoded in its cubic dependence on combinations of the $F_\phi$ and $F_\theta$ coefficients (\ref{eq:Fcoefficients1}--\ref{eq:Fcoefficients2}), which for sufficiently large $\TRS$ are directly proportional to $\TRS$. Meanwhile, the power spectrum is proportional to $(1 + \TRS^2)$. The dependence on $f_{NL}$ on $\TRS$ is then expected, in the limit where the $\TRS$ contribution dominates the $F_I$ coefficients, to be roughly $f_{NL} \propto \TRS^{-1}$. The transfer function is somewhat large ($\mathcal{O}(10^2)$) for canonical trajectory C, but this is not large enough to completely suppress the other effects. In particular, as $\ga$ is very large in this trajectory, we can see that in eq. (\ref{eq:Fcoefficients1}) especially, the transfer function term will still be subdominant compared to the early-time contribution proportional to $\ga^2 D$. Hence in this trajectory, the limit of $f_{NL} \propto \TRS^{-1}$ is not achieved due to the $F$ coefficients not being dominated by the transfer function term. Instead we see that as $F_\phi \propto \ga^2$, we have $f_{NL} \propto \ga^6 \TRS^{-4}$ which for $\ga$ also of $\mathcal{O}(10^2)$ shows that the transfer function fails to suppress the non-Gaussianity generated by the $F_I$ coefficients, which is then further amplified by the small sound speeds. $\TRS$ is similarly large in DBI trajectory A due to the extreme post-horizon-crossing evolution of the sound speeds, and in this case does suffice to suppress the less-extreme generation of non-Gaussianity from other effects, as here $\ga$ is just $\mathcal{O}(1)$, $F_I \propto \TRS$, and the usual $f_{NL} \propto \TRS^{-1}$ behaviour is realised.
\newline
\newline
None of the models studied produce either a running or a running of the running that significantly depart from the standard Planck analysis of $\beta_s = 0$ and a small $\alpha_s$ \cite{Ade:2015lrj} with all trajectories falling within $|\alpha_s|, |\beta_s| \leq O(10^{-3})$, but are in tension with the alternative analyses allowing a non-zero $\beta_s$ which support the possible existence of a large and positive running of the running \cite{Escudero2016,Cabass:2016ldu,vandeBruck:2016rfv}. While it is entirely reasonable that some other trajectory in disformally-coupled inflation is capable of generating such large runnings, given the large number of unspecified functions and parameters one could choose in this context, we find no direct evidence of this.


\subsection{Post-inflationary concerns} \label{sec:postinflation}

Of course, the value of $f_{NL}$ present at the end of inflation is not the end of the story. It would therefore be interesting to consider the post-inflationary processing of the bispectrum  due to the decays in reheating. We envisage that the decay of $\chi$ would lead to dark matter on the brane, while the decay $\phi$ would lead to radiation and eventually the standard model particles in this scenario \cite{Koivisto:2013fta}. Reheating in our model would be interesting both due to the presence of isocurvature perturbations, and the fact that the $\chi$ field's decay dynamics would be highly non-trivial due to disformal effects. To see this, consider directly adding fiducial decay terms with coupling constants $g_\phi$, $g_\chi$, to the action (\ref{eq:DisformalAction}), such as

\begin{align}
S= \frac{1}{2} \int {\rm d}^4 x \sqrt{-g} \ R  & -  \int {\rm d}^4 x \sqrt{-g}\left[\frac{1}{2} g^{\mu\nu}\phi_{,\mu}\phi_{,\nu} + U(\phi) +  g_\phi^2 \phi^2 \psi^2 \right] \nonumber \\ &-  \int {\rm d}^4 x \sqrt{-\hat{g}}\left[ \frac{1}{ 2}\hat{g}^{\mu\nu}\chi_{,\mu}\chi_{,\nu} + V(\chi) + g_\chi^2 \chi^2 \sigma^2 \right] \,.
\end{align}

When transformed to the Einstein frame as in eq. (\ref{eq:DBIModel}), one would obtain an additional term

\beq
S \supset - \int {\bd}^4 x \sqrt{-g}\left[g_\phi^2 \phi^2 \psi^2 + \frac{C^2}{\ga} g_\chi^2 \chi^2 \sigma^2 \right] = - \int {\bd}^4 x \sqrt{-g}\spar{g_\phi^2 \phi^2 \psi^2 + g_{\chi \text{,EF}}^2 \chi^2 \sigma^2} \, ,
\eeq

such that the effective coupling constant in the Einstein frame $g_{\chi \text{,EF}}^2 = C^2  g_\chi^2 / \ga$ depends on $\phi$, analogously to the modulated reheating scenario \cite{Bartolo:2004if,Bassett:2005xm,Enqvist:2012vx,Langlois:2013dh,Mazumdar:2015xka,Suyama:2007bg,Yokoyama:2011skg}, but also $\partial_\mu \phi$ (via $\ga$), complicating matters somewhat. Even in standard modulated reheating, one finds that the dynamics of the modulating field heavily affects the details of reheating \cite{Kobayashi:2013nwa}. We expect the same to be true in our case, with the difference being that the dynamics themselves are complicated by varying sound speeds and non-trivial kinetic interactions, the effects of which are largely unknown in the fast-rolling or late time oscillating limit, but expected to be non-negligible. 

Due to the non-trivial dependence of the decay rate on the field configuration and the nature of the disformal transformation, the reheating process may also vary considerably from trajectory to trajectory. For example, if the bare coupling constants $g_\phi$ and $g_\chi$ are of similar magnitude, the large value of $\ga$ achieved in Trajectory C, say, will mean the effective coupling constants in the Einstein Frame are no longer comparable as $g_{\chi \text{,EF}} \propto \ga^{-1}$ and will wildly oscillate as the $\phi$ field rolls about the minimum of its potential.

A comprehensive study of how these field dynamics affect the reheating process, and hence how much non-Gaussianity is additionally generated by it, is left to a future investigation, but one would generally expect that as in usual modulated reheating, there would be some amplification of $f_{NL}$ over the value generated purely by inflation \cite{Bernardeau:2004zz,Choi:2012te,Cicoli:2012cy}, though it is not clear whether this contribution would be sub-dominant, comparable or dominant compared to the inflationary $f_{NL}$, or even if this would be qualitatively the same for all trajectories in this model.

\section{Conclusions}\label{sec:conclusions}

In this paper we extend the results of our previous work \cite{vandeBruck:2015tna} on disformally coupled inflation to the computation of leading-order non-Gaussianities generated by the model, as well as the inclusion of a DBI kinetic term motivated by brane world realisations of this scenario. These results may be of more general interest as a first look into how non-trivial derivative couplings in multi-field models, resulting in perturbation modes propagating with non-identical sound speeds, may affect predictions and testing of non-Gaussianity. As such, the setup presented in this paper provides a motivation for such models and a playground to study the phenomenological implications of models with similar properties. 
\newline
\newline
For the trajectories and parameters we considered in this work, both fields were in the slow--roll regime, although the speed of sound for the cosmological perturbations differed significantly from 1 for a considerable amount of e--folds before the end of inflation. As such, the situation is very different from other models usually studied in the literature. That being said, for most of the trajectories considered, the non--Gaussianity parameter $f_{NL}$ remains small ($|f_{NL}|<1$), partly because it is suppressed by the late-time transfer of power from isocurvature to curvature. The exceptions to this are the trajectories of the canonical cases A and C. In C this is due to $\ga$ being much larger than in other cases, both driving the speeds of sound to near-zero and making the $F_I$ coefficients relating field perturbations to curvature very large ,such that even in the presence of a large transfer function, the non-Gaussianity remains large. In A, the sound speeds are closer to unity, but the transfer function is also smaller than in the other cases. However, this alone does not explain the difference in the amplitude of non-Gaussianity, as the transfer in cases A and B are of the same order of magnitude. As discussed above, other factors such as the size of the disformal contribution in converting field perturbations to curvature perturbations (terms containing $\ga$ or $D$ in $F_\phi \, ,F_\theta$) play a role in determining the value of $f_{NL}$.
\newline
\newline
While we have studied the slow--roll regime for both fields in this paper, which allowed us to use the formalism presented in \cite{Langlois:2008qf,Maldacena:2002vr} to calculate the non--Gaussianity, it would be interesting to check whether a fast rolling DBI--field would change the conclusions of this work. A novelity of the prototype DBI inflation was that the inflationary dynamics could be realised even in the fast-rolling regime (in these sense that the potential need not be flat when $\gamma \rightarrow \infty$) \cite{Silverstein:2003hf,Alishahiha:2004eh} but indeed the excess non-Gaussianity in this model turned out to compromise its viability. It remains to be seen whether that may be cured by the inclusion of matter upon the DBI brane.

\acknowledgments{The work of CvdB is supported by the Lancaster- Manchester-Sheffield Consortium for Fundamental Physics under STFC Grant No. ST/L000520/1. CL is supported by a STFC studentship.}

\pagebreak

\appendix

\section*{Appendices}

In these appendices, we make use of the symbol $\gad$ as defined in eq. (\ref{gammad}) to describe both canonical and DBI variants of our model, generalising the appendices in our previous paper which only described the canonical case. Below, we give expressions for the $X_n$, $Y_n$ and $Z_n$ coefficients in eqs. (\ref{eq:PEEtt}--\ref{eq:PEEss}) and the $\alpha_n$ coefficients in eq. (\ref{eq:alphas}). Note that the $\beta_n$ coefficients in eq. (\ref{eq:betas}) are unchanged by the DBI kinetic term and are hence not repeated here. Finally, we list some useful derivatives of the Lagrangian that are needed to evaluate eqs. (\ref{eq:gcoefficients}--\ref{eq:hcoefficients}).

\section{Full expressions for the $X_n$, $Y_n$ and $Z_n$ coefficients}  \label{App:XYZ}

\begin{align*}
X_1 =  & \ 2U - \gbm{2}{3} \rhochi - \ga^4 \pchi - \frac{C}{D} \rpar{\gad^3 - 3 \gad + 2} \, , \\
X_2 =  & \  U' - \frac{1}{2}\left( \left[ \gbm{2}{5}\rhochi + \ga^4 \pchi\right] \cpoc - \gbm{}{1}\left[2 \rhochi + \ga^2 \pchi \right] \dpod \right) - \frac{C}{2D} \rpar{\cpoc-\dpod} \rpar{\gad^3 - 3 \gad + 2}  \, , \\
X_3 =  & \  \left[\gad^3 + \doverc \ga^2 \left(2\rhochi + \ga^2 \pchi\right) \right]\phidot\,, \\
X_4 =  & \  \ga C^2 V'\,, \\
X_5 =  & \ \ga^3 C \chidot\,.  \\
\\
Y_1 =  & \  -\left[\gad + \doverc \rhochi \right] \phidot\,, \\
Y_2 =  & \  - \ga C \chidot\,. \\
\\
Z_1 =  & \  -2\left(\phidot^2 - U\right) - \left(\rhochi + 3\pchi \right) - \frac{C}{D} \rpar{\gad^3 - 3 \gad + 2} \,, \\
Z_2 =  & \ -U' - \frac{1}{2} \left[ \left(\rhochi - 3\pchi\right) \cpoc - \frac{\ga^2 - 1}{\ga^2} \rhochi \dpod \right] - \frac{C}{2\ga D} \rpar{\cpoc-\dpod} \rpar{\gad-1}^2  \,,\\
Z_3 =  & \  \left[\gad + \doverc \rhochi \right] \phidot\,, \\
Z_4 =  & \  -\frac{C^2 V'}{\ga}\,, \\
Z_5 =  & \  \ga C \chidot\,.
\end{align*}

\section{Full expressions for the $\al{n}$ coefficients} \label{App:AlphaDBI}
\begin{align*}
\al{1}  = & \  \gad^3 + \doverc \ga^2 \rhochi \, , \quad \al{2}  = \ 0 \, , \\
\al{3}  = & \  -\left(\gad - \doverc \ga^2 \pchi\right) \, , \quad \al{4}  = \   0 \,  , \\
\al{5}  = & \  - \phidot \left[\ga\gbp{}{3} + \doverc \ga^2 \left(\rhochi - 3 \pchi\right)\right] \, , \\
\al{6}  = & \  3 H \left[\gad^3\rpar{1 - 3 \frac{\gad^2 - 1}{\ga^2 + \ga + 1}}- \doverc \left(\ga^4 \pchi + \gbm{}{1}\left(\rhochi + \ga^2 \pchi\right)\right)\right] + \doverc \ga^2 \phidot \left[ \doverc \ga^2 \left(4 \rhochi + \ga^2 \pchi \right) \phidotdot \right.  \\
 & \left.  - \frac{1}{2}\left( \left[\gbm{4}{1} \rhochi + \gbp{}{4}\ga^2 \pchi\right] \cpoc - \left[\gbm{4}{2}\rhochi + \gbm{}{1}\ga^2 \pchi\right] \dpod \right) \right]  \\
 & + \frac{3}{2} \rpar{\cpoc-\dpod} \phidot \ga^3 \frac{1 + \gad - 2 \gad^2}{\ga^2 + \ga + 1} \, ,\\
\al{7} = & \  D \ga^3 \left(\ga^2 \phidotdot - 3 H \phidot \right)\chidot - \frac{1}{2} C \ga^3 \left(\gbp{}{1} \cpoc - \gbm{}{1} \dpod \right)\chidot \, , \\
\al{8} = & \  -\left(2 + \doverc \ga^2 \left[\gbm{4}{1} \rhochi + \ga^4 \pchi\right]\right)\phidotdot - 3 H \phidot \left(2 + \frac{\ga\rpar{\gad-1}^2\rpar{2\ga^2 + 3 \ga + 1}}{\ga^2 + \ga + 1} - \doverc \ga^2\left[\rhochi + \gbm{2}{1} \pchi\right] \right) \\
& + \frac{1}{2}\left( \left[ \left(4\ga^4 - 4\ga^2 + 2\right) \rhochi + \left(\ga^4 + 4\ga^2 - 3\right) \ga^2 \pchi\right] \cpoc - \left[\left(4\ga^4 - 5\ga^2 + 1\right) \rhochi + \gbm{}{1} \ga^4 \pchi\right] \dpod \right) \, \\
& + \frac{1}{2} \frac{C}{D} \rpar{\cpoc-\dpod} \frac{\rpar{\gad-1}^3\rpar{4 \ga^2 + 7 \ga + 4}}{\ga^2 + \ga + 1}, \\
\al{9} =  & \ U'' + \frac{1}{2}\left(\left[\gbm{}{2}\rhochi + 3\ga^2 \pchi\right] \dppod - \left[\gbm{}{1}\rhochi\right] \cppoc \right) + \frac{1}{4}\left(\left[\frac{1}{2}\gbm{4}{3}\cpoc - 2\gbm{}{1}\dpod\right]^2 \rhochi \right. \\
& \left. + \left[\gbp{}{2}\cpoc - \gbm{}{1}\dpod \right]^2 \ga^2 \pchi + \left[\frac{15}{4}\rhochi - 13 \ga^2 \pchi \right] \left(\cpoc\right)^2 \right) + \frac{\ga^2 D}{2 C} \left[\left( \left[\gbm{4}{2}\phidotdot - 3H\doverc \phidot^3\right] \rhochi \right. \right. \\
& \left. \left. + \left[\gbm{}{1}\phidotdot - 6H\phidot\right] \ga^2 \pchi\right) \dpod - \left(\left[\gbm{4}{5}\phidotdot - 3H\phidot\right] \rhochi + \left[\ga^4 \phidotdot - 3H\phidot\gbm{2}{3}\right]\pchi\right) \cpoc\right] \, \\
& + \frac{3}{2} H \phidot \rpar{\cpoc-\dpod}\frac{\ga \rpar{\gad-1}^2\rpar{2\ga^2 + 3 \ga + 1}}{\ga^2 + \ga + 1} + \frac{3}{4} \frac{C}{D} \ga^3 \rpar{\cpoc - \dpod}^2 \frac{4 \gad^2 + \gad - 2}{\gad^2 + \gad + 1} \\
&  - \frac{C}{4D} \rpar{3\gad \rpar{2 \gad^2 - 1} \rpar{\cpoc}^2 + \rpar{10\gad^2-15\gad+8} \rpar{\dpod}^2 - 2 \cpoc \dpod \rpar{8\gad^3 - 9\gad + 4}} \\
& - \frac{1}{2} \frac{C}{D} \rpar{\gad^3-3\gad+2} \rpar{\cppoc - \dppod} \, , \\
\al{10} =  & \ \left( \frac{1}{2}\left[\gbm{}{1}\dpod - \gbm{}{5} \cpoc \right] + \doverc\left[\ga^2 \phidotdot + 3H \phidot\right] \right) \ga C^2 V' \,.
\end{align*}

\section{Useful derivatives of the Lagrangian}

Out of a possible $2^4 = 16$ combinations of derivatives, only 10 of these are unique due to the symmetries $X\pc = X\cp$ and $f_{,xy} = f_{,yx}$ (standard reordering of partial derivatives). 

\begin{align}
P_{\spp\spp}\cmodel & = -\ga^3 h D \rpar{X\cc - CV} + 6 \ga^5 h^2 D (X\pc)^2 \, ,\\
& \eqbg \ga^3 h D (3 \ga^2 - 4) X\cc+ \ga^3 D^2 V \, . \\
P_{\spp\spp}\dmodel & =  h \ga^3 + P_{\spp\spp}\cmodel\, .
\end{align}

\beq
P_{\spp\scc} = P_{\scc\spp} = -\ga D
\eeq

\beq
P_{\spp\spc} = P_{\spp\scp} = P_{\spc\spp} = P_{\scp\spp} = 2 \ga^3 h D X\pc
\eeq

\beq
P_{\spc\scc} = P_{\scp\scc} = P_{\scc\spc} = P_{\scc\scp} =  0
\eeq

\beq
P_{\scc\scc} = 0
\eeq

\beq
P_{\spc\spc} = P_{\spc\scp} = P_{\scp\spc} = P_{\scp\scp} =  \ga D
\eeq

Out of a possible $2^6 = 64$ combinations of derivatives, only 10 of these are unique due to the symmetries $X\pc = X\cp$ and $f_{,xy} = f_{,yx}$ (standard reordering of partial derivatives). Of these, 6 are identically zero.

\begin{align}
P_{\spp\spp\spp}\cmodel & = -3 \ga^5 h^2 D \rpar{X\cc - CV} + 30 \ga^7 h^3 D (X\pc)^2 \, ,\\
P_{\spp\spp\spp}\dmodel & =  3 h^2 \ga^5 + P_{\spp\spp}\cmodel\, .
\end{align}

\beq
P_{\scc\scc\scc} = 0 \, .
\eeq

\begin{align}
& P_{\spc\spc\spc} = P_{\scp\spc\spc} = P_{\spc\scp\spc} = P_{\spc\spc\scp} \\ 
&= P_{\scp\scp\spc}  = P_{\scp\spc\scp} = P_{\spc\scp\scp} = P_{\scp\scp\scp} = 0 \nonumber \, .
\end{align}

\beq
P_{\spp\spp\scc} = P_{\spp\scc\spp} = P_{\scc\spp\spp} = - \ga^3 h D \, .
\eeq

\begin{align}
&P_{\spp\spp\spc} = P_{\spp\spp\scp} = P_{\spp\spc\spp}  \\ 
& = P_{\spp\scp\spp} = P_{\spc\spp\spp} = P_{\scp\spp\spp} = 6\ga^5 h^2 D X\pc \, .  \nonumber
\end{align}

\beq
P_{\scc\scc\spp} = P_{\scc\spp\scc} = P_{\spp\scc\scc} = 0 \, .
\eeq

\begin{align}
&P_{\scc\scc\spc} = P_{\scc\scc\scp} = P_{\scc\spc\scc}  \\ 
& = P_{\scc\scp\scc} = P_{\spc\scc\scc} = P_{\scp\scc\scc} =0 \, . \nonumber
\end{align}

\begin{align}
& P_{\spp\spc\spc} = P_{\spp\scp\spc} = P_{\spp\spc\scp} = P_{\spp\scp\scp} \\ 
& = P_{\spc\spp\spc} = P_{\scp\spp\spc} = P_{\spc\spp\scp} = P_{\scp\spp\scp}   \nonumber \\ 
&= P_{\spc\spc\spp}  = P_{\scp\spc\spp} = P_{\spc\scp\spp} = P_{\scp\scp\spp} = \ga^3 h D \nonumber \, .
\end{align}

\begin{align}
& P_{\scc\spc\spc} = P_{\scc\scp\spc} = P_{\scc\spc\scp} = P_{\scc\scp\scp} \\ 
& = P_{\spc\scc\spc} = P_{\scp\scc\spc} = P_{\spc\scc\scp} = P_{\scp\scc\scp}   \nonumber \\ 
&= P_{\spc\spc\scc}  = P_{\scp\spc\scc} = P_{\spc\scp\scc} = P_{\scp\scp\scc} = 0\nonumber \, .
\end{align}

\begin{align}
& P_{\spp\scc\spc} = P_{\spp\scc\scp} = P_{\scc\spp\spc} = P_{\scc\spp\scp} \\ 
& = P_{\spp\spc\scc} = P_{\spp\scp\scc} = P_{\scc\spc\spp} = P_{\scc\scp\spp}   \nonumber \\ 
&= P_{\spc\spp\scc}  = P_{\scp\spp\scc} = P_{\spc\scc\spp} = P_{\scp\scc\spp} = 0 \nonumber \, .
\end{align}

\bibliography{DBING_bib}

\providecommand{\href}[2]{#2}\begingroup\raggedright\begin{thebibliography}{10}

\bibitem{Will:2014kxa}
C.~M. Will, {\it {The Confrontation between General Relativity and
  Experiment}},  {\em Living Rev. Rel.} {\bf 17} (2014) 4,
  [\href{http://arxiv.org/abs/1403.7377}{{\tt arXiv:1403.7377}}].

\bibitem{Abbott:2016blz}
{\bf Virgo, LIGO Scientific} Collaboration, B.~P. Abbott et~al., {\it
  {Observation of Gravitational Waves from a Binary Black Hole Merger}},  {\em
  Phys. Rev. Lett.} {\bf 116} (2016), no.~6 061102,
  [\href{http://arxiv.org/abs/1602.03837}{{\tt arXiv:1602.03837}}].

\bibitem{Clifton:2011jh}
T.~Clifton, P.~G. Ferreira, A.~Padilla, and C.~Skordis, {\it {Modified Gravity
  and Cosmology}},  {\em Phys. Rept.} {\bf 513} (2012) 1--189,
  [\href{http://arxiv.org/abs/1106.2476}{{\tt arXiv:1106.2476}}].

\bibitem{Bekenstein:1992pj}
J.~D. Bekenstein, {\it {The Relation between physical and gravitational
  geometry}},  {\em Phys. Rev.} {\bf D48} (1993) 3641--3647,
  [\href{http://arxiv.org/abs/gr-qc/9211017}{{\tt gr-qc/9211017}}].

\bibitem{Kaloper:2003yf}
N.~Kaloper, {\it {Disformal inflation}},  {\em Phys. Lett.} {\bf B583} (2004)
  1--13, [\href{http://arxiv.org/abs/hep-ph/0312002}{{\tt hep-ph/0312002}}].

\bibitem{Zumalacarregui:2010wj}
M.~Zumalacarregui, T.~S. Koivisto, D.~F. Mota, and P.~Ruiz-Lapuente, {\it
  {Disformal Scalar Fields and the Dark Sector of the Universe}},  {\em JCAP}
  {\bf 1005} (2010) 038, [\href{http://arxiv.org/abs/1004.2684}{{\tt
  arXiv:1004.2684}}].

\bibitem{Sakstein:2015jca}
J.~Sakstein and S.~Verner, {\it {Disformal Gravity Theories: A Jordan Frame
  Analysis}},  \href{http://arxiv.org/abs/1509.05679}{{\tt arXiv:1509.05679}}.

\bibitem{Sakstein:2014aca}
J.~Sakstein, {\it {Towards Viable Cosmological Models of Disformal Theories of
  Gravity}},  {\em Phys. Rev.} {\bf D91} (2015), no.~2 024036,
  [\href{http://arxiv.org/abs/1409.7296}{{\tt arXiv:1409.7296}}].

\bibitem{Emond:2015efw}
W.~T. Emond and P.~M. Saffin, {\it {Disformally self-tuning gravity}},  {\em
  JHEP} {\bf 03} (2016) 161, [\href{http://arxiv.org/abs/1511.02055}{{\tt
  arXiv:1511.02055}}].

\bibitem{Burrage:2016myt}
C.~Burrage, S.~Cespedes, and A.-C. Davis, {\it {Disformal transformations on
  the CMB}},  \href{http://arxiv.org/abs/1604.08038}{{\tt arXiv:1604.08038}}.

\bibitem{Brax:2015hma}
P.~Brax, C.~Burrage, and C.~Englert, {\it {Disformal dark energy at
  colliders}},  \href{http://arxiv.org/abs/1506.04057}{{\tt arXiv:1506.04057}}.

\bibitem{Minamitsuji:2016hkk}
M.~Minamitsuji and H.~O. Silva, {\it {Relativistic stars in scalar-tensor
  theories with disformal coupling}},  {\em Phys. Rev.} {\bf D93} (2016),
  no.~12 124041, [\href{http://arxiv.org/abs/1604.07742}{{\tt
  arXiv:1604.07742}}].

\bibitem{Bettoni:2016mij}
D.~Bettoni, J.~M. Ezquiaga, K.~Hinterbichler, and M.~Zumalacárregui, {\it
  {Gravitational Waves and the Fate of Scalar-Tensor Gravity}},
  \href{http://arxiv.org/abs/1608.01982}{{\tt arXiv:1608.01982}}.

\bibitem{Zumalacarregui:2012us}
M.~Zumalacarregui, T.~S. Koivisto, and D.~F. Mota, {\it {DBI Galileons in the
  Einstein Frame: Local Gravity and Cosmology}},  {\em Phys. Rev.} {\bf D87}
  (2013) 083010, [\href{http://arxiv.org/abs/1210.8016}{{\tt
  arXiv:1210.8016}}].

\bibitem{Zumalacarregui:2013pma}
M.~Zumalacárregui and J.~García-Bellido, {\it {Transforming gravity: from
  derivative couplings to matter to second-order scalar-tensor theories beyond
  the Horndeski Lagrangian}},  {\em Phys. Rev.} {\bf D89} (2014) 064046,
  [\href{http://arxiv.org/abs/1308.4685}{{\tt arXiv:1308.4685}}].

\bibitem{Bettoni:2013diz}
D.~Bettoni and S.~Liberati, {\it {Disformal invariance of second order
  scalar-tensor theories: Framing the Horndeski action}},  {\em Phys. Rev.}
  {\bf D88} (2013), no.~8 084020, [\href{http://arxiv.org/abs/1306.6724}{{\tt
  arXiv:1306.6724}}].

\bibitem{Arroja:2015wpa}
F.~Arroja, N.~Bartolo, P.~Karmakar, and S.~Matarrese, {\it {The two faces of
  mimetic Horndeski gravity: disformal transformations and Lagrange
  multiplier}},  {\em JCAP} {\bf 1509} (2015) 051,
  [\href{http://arxiv.org/abs/1506.08575}{{\tt arXiv:1506.08575}}].

\bibitem{Achour:2016rkg}
J.~Ben~Achour, D.~Langlois, and K.~Noui, {\it {Degenerate higher order
  scalar-tensor theories beyond Horndeski and disformal transformations}},
  {\em Phys. Rev.} {\bf D93} (2016), no.~12 124005,
  [\href{http://arxiv.org/abs/1602.08398}{{\tt arXiv:1602.08398}}].

\bibitem{vandeBruck:2012vq}
C.~van~de Bruck and G.~Sculthorpe, {\it {Modified Gravity and the Radiation
  Dominated Epoch}},  {\em Phys. Rev.} {\bf D87} (2013), no.~4 044004,
  [\href{http://arxiv.org/abs/1210.2168}{{\tt arXiv:1210.2168}}].

\bibitem{vandeBruck:2015ida}
C.~van~de Bruck and J.~Morrice, {\it {Disformal couplings and the dark sector
  of the universe}},  {\em JCAP} {\bf 1504} (2015) 036,
  [\href{http://arxiv.org/abs/1501.03073}{{\tt arXiv:1501.03073}}].

\bibitem{Motohashi:2015pra}
H.~Motohashi and J.~White, {\it {Disformal invariance of curvature
  perturbation}},  \href{http://arxiv.org/abs/1504.00846}{{\tt
  arXiv:1504.00846}}.

\bibitem{Minamitsuji:2014waa}
M.~Minamitsuji, {\it {Disformal transformation of cosmological perturbations}},
   {\em Phys. Lett.} {\bf B737} (2014) 139--150,
  [\href{http://arxiv.org/abs/1409.1566}{{\tt arXiv:1409.1566}}].

\bibitem{Domenech:2015hka}
G.~Domènech, A.~Naruko, and M.~Sasaki, {\it {Cosmological disformal
  invariance}},  \href{http://arxiv.org/abs/1505.00174}{{\tt
  arXiv:1505.00174}}.

\bibitem{Hagala:2015paa}
R.~Hagala, C.~Llinares, and D.~F. Mota, {\it {Cosmological simulations with
  disformally coupled symmetron fields}},
  \href{http://arxiv.org/abs/1504.07142}{{\tt arXiv:1504.07142}}.

\bibitem{Yuan:2015tta}
F.-F. Yuan and P.~Huang, {\it {Induced geometry from disformal
  transformation}},  {\em Phys. Lett.} {\bf B744} (2015) 120--124,
  [\href{http://arxiv.org/abs/1501.06135}{{\tt arXiv:1501.06135}}].

\bibitem{Bittencourt:2015ypa}
E.~Bittencourt, I.~P. Lobo, and G.~G. Carvalho, {\it {On the disformal
  invariance of the Dirac equation}},  {\em Class. Quant. Grav.} {\bf 32}
  (2015) 185016, [\href{http://arxiv.org/abs/1505.03415}{{\tt
  arXiv:1505.03415}}].

\bibitem{Huang:2015hja}
P.~Huang and F.-F. Yuan, {\it {Disformal transformation in Newton-Cartan
  geometry}},  {\em Eur. Phys. J.} {\bf C76} (2016), no.~8 436,
  [\href{http://arxiv.org/abs/1509.06005}{{\tt arXiv:1509.06005}}].

\bibitem{Carvalho:2015omv}
G.~G. Carvalho, I.~P. Lobo, and E.~Bittencourt, {\it {Extended disformal
  approach in the scenario of Rainbow Gravity}},  {\em Phys. Rev.} {\bf D93}
  (2016), no.~4 044005, [\href{http://arxiv.org/abs/1511.00495}{{\tt
  arXiv:1511.00495}}].

\bibitem{Bittencourt:2016smd}
E.~Bittencourt, U.~Moschella, M.~Novello, and J.~D. Toniato, {\it {More about
  scalar gravity}},  {\em Phys. Rev.} {\bf D93} (2016), no.~12 124023,
  [\href{http://arxiv.org/abs/1605.09778}{{\tt arXiv:1605.09778}}].

\bibitem{vandeBruck:2015tna}
C.~van~de Bruck, T.~Koivisto, and C.~Longden, {\it {Disformally coupled
  inflation}},  {\em JCAP} {\bf 1603} (2016), no.~03 006,
  [\href{http://arxiv.org/abs/1510.01650}{{\tt arXiv:1510.01650}}].

\bibitem{Bettoni:2015wta}
D.~Bettoni and M.~Zumalacárregui, {\it {Kinetic mixing in scalar-tensor
  theories of gravity}},  {\em Phys. Rev.} {\bf D91} (2015) 104009,
  [\href{http://arxiv.org/abs/1502.02666}{{\tt arXiv:1502.02666}}].

\bibitem{Silverstein:2003hf}
E.~Silverstein and D.~Tong, {\it {Scalar speed limits and cosmology:
  Acceleration from D-cceleration}},  {\em Phys. Rev.} {\bf D70} (2004) 103505,
  [\href{http://arxiv.org/abs/hep-th/0310221}{{\tt hep-th/0310221}}].

\bibitem{Alishahiha:2004eh}
M.~Alishahiha, E.~Silverstein, and D.~Tong, {\it {DBI in the sky}},  {\em Phys.
  Rev.} {\bf D70} (2004) 123505,
  [\href{http://arxiv.org/abs/hep-th/0404084}{{\tt hep-th/0404084}}].

\bibitem{Baumann:2010sx}
D.~Baumann, A.~Dymarsky, S.~Kachru, I.~R. Klebanov, and L.~McAllister, {\it
  {D3-brane Potentials from Fluxes in AdS/CFT}},  {\em JHEP} {\bf 06} (2010)
  072, [\href{http://arxiv.org/abs/1001.5028}{{\tt arXiv:1001.5028}}].

\bibitem{Cembranos:2003mr}
J.~A.~R. Cembranos, A.~Dobado, and A.~L. Maroto, {\it {Brane world dark
  matter}},  {\em Phys. Rev. Lett.} {\bf 90} (2003) 241301,
  [\href{http://arxiv.org/abs/hep-ph/0302041}{{\tt hep-ph/0302041}}].

\bibitem{Koivisto:2013fta}
T.~Koivisto, D.~Wills, and I.~Zavala, {\it {Dark D-brane Cosmology}},  {\em
  JCAP} {\bf 1406} (2014) 036, [\href{http://arxiv.org/abs/1312.2597}{{\tt
  arXiv:1312.2597}}].

\bibitem{Koivisto:2014gia}
T.~S. Koivisto and F.~R. Urban, {\it {Disformal vectors and anisotropies on a
  warped brane\protect Hulluilla on Halvat Huvit}},  {\em JCAP} {\bf 1503}
  (2015), no.~03 003, [\href{http://arxiv.org/abs/1407.3445}{{\tt
  arXiv:1407.3445}}].

\bibitem{Cembranos:2016jun}
J.~A.~R. Cembranos and A.~L. Maroto, {\it {Disformal scalars as dark matter
  candidates: Branon phenomenology}},  {\em Int. J. Mod. Phys.} {\bf 31}
  (2016), no.~14n15 1630015, [\href{http://arxiv.org/abs/1602.07270}{{\tt
  arXiv:1602.07270}}].

\bibitem{Martin:2013tda}
J.~Martin, C.~Ringeval, and V.~Vennin, {\it {Encyclop\ae dia Inflationaris}},
  {\em Phys. Dark Univ.} {\bf 5-6} (2014) 75--235,
  [\href{http://arxiv.org/abs/1303.3787}{{\tt arXiv:1303.3787}}].

\bibitem{Escudero2016}
M.~Escudero, H.~Ramírez, L.~Boubekeur, E.~Giusarma, and O.~Mena, {\it The
  present and future of the most favoured inflationary models after planck
  2015},  {\em Journal of Cosmology and Astroparticle Physics} {\bf 2016}
  (2016), no.~02 020.

\bibitem{Ade:2015lrj}
{\bf Planck} Collaboration, P.~A.~R. Ade et~al., {\it {Planck 2015 results. XX.
  Constraints on inflation}},  \href{http://arxiv.org/abs/1502.02114}{{\tt
  arXiv:1502.02114}}.

\bibitem{Ade:2015tva}
{\bf BICEP2, Planck} Collaboration, P.~A.~R. Ade et~al., {\it {Joint Analysis
  of BICEP2/$Keck Array$ and $Planck$ Data}},  {\em Phys. Rev. Lett.} {\bf 114}
  (2015) 101301, [\href{http://arxiv.org/abs/1502.00612}{{\tt
  arXiv:1502.00612}}].

\bibitem{Seery:2005wm}
D.~Seery and J.~E. Lidsey, {\it {Primordial non-Gaussianities in single field
  inflation}},  {\em JCAP} {\bf 0506} (2005) 003,
  [\href{http://arxiv.org/abs/astro-ph/0503692}{{\tt astro-ph/0503692}}].

\bibitem{Peterson2011}
C.~M. Peterson and M.~Tegmark, {\it Non-gaussianity in two-field inflation},
  {\em Phys. Rev. D} {\bf 84} (Jul, 2011) 023520.

\bibitem{Maldacena:2002vr}
J.~M. Maldacena, {\it {Non-Gaussian features of primordial fluctuations in
  single field inflationary models}},  {\em JHEP} {\bf 05} (2003) 013,
  [\href{http://arxiv.org/abs/astro-ph/0210603}{{\tt astro-ph/0210603}}].

\bibitem{Langlois:2008qf}
D.~Langlois, S.~Renaux-Petel, D.~A. Steer, and T.~Tanaka, {\it {Primordial
  perturbations and non-Gaussianities in DBI and general multi-field
  inflation}},  {\em Phys. Rev.} {\bf D78} (2008) 063523,
  [\href{http://arxiv.org/abs/0806.0336}{{\tt arXiv:0806.0336}}].

\bibitem{Arroja:2008yy}
F.~Arroja, S.~Mizuno, and K.~Koyama, {\it {Non-gaussianity from the bispectrum
  in general multiple field inflation}},  {\em JCAP} {\bf 0808} (2008) 015,
  [\href{http://arxiv.org/abs/0806.0619}{{\tt arXiv:0806.0619}}].

\bibitem{Palma:2015eth}
G.~A. {Palma}, {\it {Untangling features in the primordial spectra}},  {\em
  JCAP} {\bf 4} (Apr., 2015) 35, [\href{http://arxiv.org/abs/1412.5615}{{\tt
  arXiv:1412.5615}}].

\bibitem{Bartolo:2004if}
N.~Bartolo, E.~Komatsu, S.~Matarrese, and A.~Riotto, {\it {Non-Gaussianity from
  inflation: Theory and observations}},  {\em Phys. Rept.} {\bf 402} (2004)
  103--266, [\href{http://arxiv.org/abs/astro-ph/0406398}{{\tt
  astro-ph/0406398}}].

\bibitem{Byrnes:2010em}
C.~T. Byrnes and K.-Y. Choi, {\it {Review of local non-Gaussianity from
  multi-field inflation}},  {\em Adv. Astron.} {\bf 2010} (2010) 724525,
  [\href{http://arxiv.org/abs/1002.3110}{{\tt arXiv:1002.3110}}].

\bibitem{Chen:2010xka}
X.~Chen, {\it {Primordial Non-Gaussianities from Inflation Models}},  {\em Adv.
  Astron.} {\bf 2010} (2010) 638979,
  [\href{http://arxiv.org/abs/1002.1416}{{\tt arXiv:1002.1416}}].

\bibitem{Chen:2006nt}
X.~Chen, M.-x. Huang, S.~Kachru, and G.~Shiu, {\it {Observational signatures
  and non-Gaussianities of general single field inflation}},  {\em JCAP} {\bf
  0701} (2007) 002, [\href{http://arxiv.org/abs/hep-th/0605045}{{\tt
  hep-th/0605045}}].

\bibitem{Huang:2007hh}
M.-x. Huang, G.~Shiu, and B.~Underwood, {\it {Multifield DBI Inflation and
  Non-Gaussianities}},  {\em Phys. Rev.} {\bf D77} (2008) 023511,
  [\href{http://arxiv.org/abs/0709.3299}{{\tt arXiv:0709.3299}}].

\bibitem{Cai:2008if}
Y.-F. Cai and W.~Xue, {\it {N-flation from multiple DBI type actions}},  {\em
  Phys. Lett.} {\bf B680} (2009) 395--398,
  [\href{http://arxiv.org/abs/0809.4134}{{\tt arXiv:0809.4134}}].

\bibitem{Cai:2009hw}
Y.-F. Cai and H.-Y. Xia, {\it {Inflation with multiple sound speeds: a model of
  multiple DBI type actions and non-Gaussianities}},  {\em Phys. Lett.} {\bf
  B677} (2009) 226--234, [\href{http://arxiv.org/abs/0904.0062}{{\tt
  arXiv:0904.0062}}].

\bibitem{Emery:2012sm}
J.~Emery, G.~Tasinato, and D.~Wands, {\it {Local non-Gaussianity from rapidly
  varying sound speeds}},  {\em JCAP} {\bf 1208} (2012) 005,
  [\href{http://arxiv.org/abs/1203.6625}{{\tt arXiv:1203.6625}}].

\bibitem{Emery:2013yua}
J.~Emery, G.~Tasinato, and D.~Wands, {\it {Mixed non-Gaussianity in
  multiple-DBI inflation}},  {\em JCAP} {\bf 1305} (2013) 021,
  [\href{http://arxiv.org/abs/1303.3975}{{\tt arXiv:1303.3975}}].

\bibitem{Pi:2011tv}
S.~Pi and D.~Wang, {\it {Cosmological perturbations in inflation with multiple
  sound speeds}},  {\em Nucl. Phys.} {\bf B862} (2012) 409--429,
  [\href{http://arxiv.org/abs/1107.0813}{{\tt arXiv:1107.0813}}].

\bibitem{Tsujikawa:2003ccc}
S.~{Tsujikawa}, D.~{Parkinson}, and B.~A. {Bassett}, {\it
  {Correlation-consistency cartography of the double-inflation landscape}},
  {\em Phys. Rev. D} {\bf 67} (Apr., 2003) 083516,
  [\href{http://arxiv.org/abs/astro-ph/0210322}{{\tt astro-ph/0210322}}].

\bibitem{Langlois:2008mn}
D.~Langlois and S.~Renaux-Petel, {\it {Perturbations in generalized multi-field
  inflation}},  {\em JCAP} {\bf 0804} (2008) 017,
  [\href{http://arxiv.org/abs/0801.1085}{{\tt arXiv:0801.1085}}].

\bibitem{Klebanov:2000hb}
I.~R. Klebanov and M.~J. Strassler, {\it {Supergravity and a confining gauge
  theory: Duality cascades and chi SB resolution of naked singularities}},
  {\em JHEP} {\bf 08} (2000) 052,
  [\href{http://arxiv.org/abs/hep-th/0007191}{{\tt hep-th/0007191}}].

\bibitem{Wands:2002bn}
D.~Wands, N.~Bartolo, S.~Matarrese, and A.~Riotto, {\it {An Observational test
  of two-field inflation}},  {\em Phys. Rev.} {\bf D66} (2002) 043520,
  [\href{http://arxiv.org/abs/astro-ph/0205253}{{\tt astro-ph/0205253}}].

\bibitem{DiMarco:2002eb}
F.~Di~Marco, F.~Finelli, and R.~Brandenberger, {\it {Adiabatic and isocurvature
  perturbations for multifield generalized Einstein models}},  {\em Phys. Rev.}
  {\bf D67} (2003) 063512, [\href{http://arxiv.org/abs/astro-ph/0211276}{{\tt
  astro-ph/0211276}}].

\bibitem{Longden:2016fgu}
C.~Longden, {\it {The adiabatic/entropy decomposition in $P(\phi^I,X^{IJ})$
  theories with multiple sound speeds}},
  \href{http://arxiv.org/abs/1611.03481}{{\tt arXiv:1611.03481}}.

\bibitem{Cabass:2016ldu}
G.~Cabass, E.~Di~Valentino, A.~Melchiorri, E.~Pajer, and J.~Silk, {\it
  {Constraints on the running of the running of the scalar tilt from CMB
  anisotropies and spectral distortions}},  {\em Phys. Rev.} {\bf D94} (2016),
  no.~2 023523, [\href{http://arxiv.org/abs/1605.00209}{{\tt
  arXiv:1605.00209}}].

\bibitem{vandeBruck:2016rfv}
C.~van~de Bruck and C.~Longden, {\it {Running of the Running and Entropy
  Perturbations During Inflation}},  {\em Phys. Rev.} {\bf D94} (2016), no.~2
  021301, [\href{http://arxiv.org/abs/1606.02176}{{\tt arXiv:1606.02176}}].

\bibitem{Bassett:2005xm}
B.~A. Bassett, S.~Tsujikawa, and D.~Wands, {\it {Inflation dynamics and
  reheating}},  {\em Rev. Mod. Phys.} {\bf 78} (2006) 537--589,
  [\href{http://arxiv.org/abs/astro-ph/0507632}{{\tt astro-ph/0507632}}].

\bibitem{Enqvist:2012vx}
K.~Enqvist and S.~Rusak, {\it {Modulated preheating and isocurvature
  perturbations}},  {\em JCAP} {\bf 1303} (2013) 017,
  [\href{http://arxiv.org/abs/1210.2192}{{\tt arXiv:1210.2192}}].

\bibitem{Langlois:2013dh}
D.~Langlois and T.~Takahashi, {\it {Density Perturbations from Modulated Decay
  of the Curvaton}},  {\em JCAP} {\bf 1304} (2013) 014,
  [\href{http://arxiv.org/abs/1301.3319}{{\tt arXiv:1301.3319}}].

\bibitem{Mazumdar:2015xka}
A.~Mazumdar and K.~P. Modak, {\it {Constraints on variations in inflaton decay
  rate from modulated preheating}},  {\em JCAP} {\bf 1606} (2016), no.~06 030,
  [\href{http://arxiv.org/abs/1506.01469}{{\tt arXiv:1506.01469}}].

\bibitem{Suyama:2007bg}
T.~Suyama and M.~Yamaguchi, {\it {Non-Gaussianity in the modulated reheating
  scenario}},  {\em Phys. Rev.} {\bf D77} (2008) 023505,
  [\href{http://arxiv.org/abs/0709.2545}{{\tt arXiv:0709.2545}}].

\bibitem{Yokoyama:2011skg}
S.~Yokoyama, K.~Kamada, and K.~Kohri, {\it {Iso-curvature fluctuations in
  modulated reheating scenario}},  in {\em {Proceedings, 20th Workshop on
  General Relativity and Gravitation in Japan (JGRG20): Kyoto, Japan, September
  21-25, 2010}}, pp.~456--459, 2011.

\bibitem{Kobayashi:2013nwa}
N.~Kobayashi, T.~Kobayashi, and A.~L. Erickcek, {\it {Rolling in the Modulated
  Reheating Scenario}},  {\em JCAP} {\bf 1401} (2014) 036,
  [\href{http://arxiv.org/abs/1308.4154}{{\tt arXiv:1308.4154}}].

\bibitem{Bernardeau:2004zz}
F.~Bernardeau, L.~Kofman, and J.-P. Uzan, {\it {Modulated fluctuations from
  hybrid inflation}},  {\em Phys. Rev.} {\bf D70} (2004) 083004,
  [\href{http://arxiv.org/abs/astro-ph/0403315}{{\tt astro-ph/0403315}}].

\bibitem{Choi:2012te}
K.-Y. Choi and O.~Seto, {\it {Modulated reheating by curvaton}},  {\em Phys.
  Rev.} {\bf D85} (2012) 123528, [\href{http://arxiv.org/abs/1204.1419}{{\tt
  arXiv:1204.1419}}]. [Erratum: Phys. Rev.D87,no.2,029902(2013)].

\bibitem{Cicoli:2012cy}
M.~Cicoli, G.~Tasinato, I.~Zavala, C.~P. Burgess, and F.~Quevedo, {\it
  {Modulated Reheating and Large Non-Gaussianity in String Cosmology}},  {\em
  JCAP} {\bf 1205} (2012) 039, [\href{http://arxiv.org/abs/1202.4580}{{\tt
  arXiv:1202.4580}}].

\end{thebibliography}\endgroup

\end{document}